\documentclass[prd,aps,eqsecnum]{revtex4}
\usepackage{epsfig}
\usepackage{amsfonts}
\usepackage{rotating} 
\usepackage{sparticles}
\usepackage{dcolumn}
\usepackage{bm}
\usepackage{hyperref}
\hypersetup{
    bookmarks=true,         
    unicode=false,          
    pdftoolbar=true,        
    pdfmenubar=true,        
    pdffitwindow=false,     
    pdfstartview={FitH},    
    pdftitle={My title},    
    pdfauthor={Author},     
    pdfsubject={Subject},   
    pdfcreator={Creator},   
    pdfproducer={Producer}, 
    pdfkeywords={keyword1} {key2} {key3}, 
    pdfnewwindow=true,      
    colorlinks=true,       
    linkcolor=red,          
    citecolor=cyan,        
    filecolor=magenta,      
    urlcolor=green,           
    linktocpage=true
}
\usepackage{amsmath,amssymb,amsthm,graphicx,latexsym}
\usepackage{subfigure}
\usepackage{pstricks}
\usepackage{color}

\usepackage{mathrsfs}

\newcommand{\be}{\begin{equation}}
\newcommand{\ee}{\end{equation}}
\newcommand{\bea}{\begin{eqnarray}}
\newcommand{\eea}{\end{eqnarray}}
\newcommand{\bml}{\begin{subequations}}
\newcommand{\eml}{\end{subequations}}
\newcommand{\bfig}{\begin{figure}}
\newcommand{\efig}{\end{figure}}

\begin{document}

\title{ Thermodynamics of Charged Kalb Ramond AdS black hole in presence of Gauss-Bonnet coupling}

\author{{\textcolor{black}{Sayantan Choudhury}}$^{1}$\footnote{Electronic address: {sayanphysicsisi@gmail.com}} ${}^{}$
and {\textcolor{black}{Soumitra SenGupta}}$^{2}$
\footnote{Electronic address: {tpssg@iacs.res.in
}} ${}^{}$}
\affiliation{$^1$Physics and Applied Mathematics Unit, Indian Statistical Institute, 203 B.T. Road, Kolkata 700 108, India\\
$^2$Department of Theoretical Physics,
Indian Association for the Cultivation of Science,
2A and 2B Raja S.C. Mullick Road,
Kolkata - 700 032, India.
}

\date{\today}
\begin{abstract}
We study the role of the Gauss-Bonnet corrections to the gravity
action on the charged AdS black hole in presence of rank 3 antisymmetric Kalb Ramond tensor field strength.
Analyzing the branch singularity and the killing horizon,
we explicitly derive various thermodynamic parameters and study their behaviour in presence of five dimensional Gauss-Bonnet coupling in AdS space-time.
The possibility of a second order phase transition is explored in the light of {\bf AdS/CMT} correspondence and various critical 
exponents associated with the discontinuities of the various thermodynamic parameters are determined.
We further comment on the universality of the well known {\it Rushbrooke Josephson scaling law} and derive  a relation between the 
degree of homogeneity appearing in various free energies and the critical exponents by homogeneous hypothesis test.
By making use of the constraints appearing from Hawking temperature and Gauss-Bonnet extended gravity version of 
{\it Kubo formula} we introduce a bound on
the five dimensional Gauss-Bonnet coupling and the viscosity entropy ratio in the four dimensional holographic {\it Conformal Field Theory} ({\bf CFT})
dual. This yields a fractional deviation in viscosity entropy ratio from the result obtained from Einstein gravity.
\end{abstract}


\maketitle
\tableofcontents
\section{\bf Introduction}
\label{intro}

Different features of Einstein's gravity in the realm of $(3 + 1)$ space-time dimensions have been studied through decades.
If we embed such gravity on higher dimensional anti-deSiiter (AdS) manifold then the theory becomes
 non-renormalizable \cite{barvi,max,dami} which is obviously 
a serious problem. The leading order quantum gravity corrections in a  higher dimensional bulk manifold have been studied specially in the context of string theory. 
String theory is one of the realization where the two loop correction on the CFT disk amplitude via
 the inverse of Regge slope (or string tension) gives Gauss-Bonnet (GB) correction \cite{sayan1,sayan2,sayan3,gasp} to the 
usual Einstein-Hilbert action in its effective field theory version (below {\bf UV} cutoff). Since GB correction is quadratic topological invariant 
in four dimension, it will always contribute in the dimension $D\geq 5$. On the other hand CFT \cite{ginspa,matt,carli,kach,witten,maldacena} is realized in the boundary of the 
prescribed AdS bulk topological manifold \cite{cardy,yu,sen1,serge}. Most importantly a perfect one to one mapping between bulk and boundary parameters can only be realized iff
the dimensionality of the bulk AdS manifold is $D>4$ and the corresponding boundary dual CFT is embedded on $D=4$. This clearly suggests that the unification
 between the quantum gravity correction and the correspondence can only be realized at least in ${\bf AdS_{5}/CFT_{4}}$ theory
 \cite{igor,malda2,witten2,witten3,mathur,witten4,albion,witten5} which is our present focus.

 In this article we start with a five dimensional bulk manifold where GB correction are included and the 
Kalb Ramond rank three antisymmetric tensor field is embedded on ${\bf AdS_{5}}$ where the corresponding extra dimension is non-compact.  
We also consider a single localized brane boundary on which dual holographic CFT can be clearly visualized. 
We then make  a comprehensive study of AdS black hole thermodynamics and equilibrium statistical mechanics 
 and its implications on phase transition and ${\bf AdS_{5}/CFT_{4}}$ correspondence. 
We determine the physically acceptable metric function from the solution of Einstein's equation in presence of GB correction,
 its asymptotic behaviour and study of branch singularity and killing horizon.
Hence we study different thermodynamic parameters to examine their behaviour 
in the context of black hole thermodynamics. We also study {\bf AdS/CMT} correspondence \cite{burg,andy,dori,lin,chow1,subir,subir2,subir3,subir4} by determining
 the values of the critical exponents associated with the discontinuities
in various thermodynamic parameters and the corresponding order of the phase transition in AdS space-time.
 Then we make a comment on the validity and universality of {\it Rushbrooke Josephson scaling laws} \cite{golden,ma} commonly used in Condensed Matter theory (CMT).
 After that we establish the connection between the degree of homogeneity in free energy with the critical exponents by homogeneous hypothesis testing method.
 Further, we  study ${\bf AdS_{5}/CFT_{4}}$ by determining the relation between five dimensional Gauss-Bonnet coupling ($\alpha_{(5)}$) with the well known $\frac{\eta}{s}$ ratio
appearing in the 4D CFT holographic dual theory. We also estimate the numerical bound on  $\frac{\eta}{s}$ ratio by fixing the lower cutoff and upper
 cutoff of $\alpha_{(5)}$ from the thermodynamical behaviour in the bulk theory.

\section{\bf Einstein Gauss-Bonnet model with Kalb Ramond field in a 5-dimensional bulk spacetime}
\label{model}
We start our discussion with a model 
on a warped product manifold with an  extra dimension in a single brane set up \cite{meda,sahani}.
In this five dimensional framework the model is described by the following action:
\be\begin{array}{llllll}\label{action}
   \displaystyle  S_{(5)}=S_{EH}+S_{GB}+S_{Bulk}+S_{Brane},
   \end{array}\ee
where the contribution from the gravity sector is given by the {\it Einstein-Hilbert}, {\it Gauss-Bonnet} in the bulk geometry such that,
\be\begin{array}{llllll}\label{eh}
   \displaystyle  S_{EH}=\frac{M^{3}_{(5)}}{2}\int d^{5}x \sqrt{-g_{(5)}}R_{(5)},
   \end{array}\ee

\be\begin{array}{llllll}\label{gb}
   \displaystyle  S_{GB}=\frac{\alpha_{(5)}M_{(5)}}{2}\int d^{5}x \sqrt{-g_{(5)}}\left[R^{ABCD(5)}R^{(5)}_{ABCD}-4R^{AB(5)}R^{(5)}_{AB}+R^{2}_{(5)}\right],
   \end{array}\ee

with $A,B,C,D=0(\Rightarrow t),1(\Rightarrow r),2(\Rightarrow x),3(\Rightarrow y),4(\Rightarrow z)$. It is important to mention here that the extra
 dimension $r$ is non-compact.
Other contributions come from bulk rank 3 antisymmetric tensor Kalb Ramond field and single brane sector are  given as:
\be\begin{array}{llllll}\label{bulk}
   \displaystyle  S_{Bulk}=\int d^{5}x \sqrt{-g_{(5)}}\left[-M^{3}_{(5)}{\cal H}_{ABC}(r,x_{\mu}){\cal H}^{ABC}(r,x_{\mu})-2\Lambda_{(5)}\right],
   \end{array}\ee

\be\begin{array}{llllll}\label{brane1}
   \displaystyle  S_{Brane}=\int d^{4}\xi \sqrt{- h_{(4)}}\left[{\cal L}^{field}-T_{(3)}\right].
   \end{array}\ee
 Throughout 
the article we use $\alpha_{(5)}$ as Gauss-Bonnet coupling, $d^{4}\xi$ is the four dimensional counterpart of the five dimensional world volume $d^{4}x$ and
$h_{(4)}$ is the determinant of the four dimensional induced metric . 
In equation(\ref{brane1}) ${\cal L}^{field}$ represent brane Lagrangian which contains brane fields and $T_{(3)}$ be the brane tension for the single brane. 

The background five dimensional metric describing slice of the warped product manifold ${{\cal M}^{2}\otimes {\cal K}^{3}}$ in the spacelike hypersurface is given by \cite{meda,sahani},
\be\begin{array}{llllll}\label{brane}
   \displaystyle ds^{2}_{(5)}=g_{AB}dx^{A}dx^{B}=-h(r)dt^{2}+f(r)dr^{2}+r^{2}\gamma_{ij}dx^{i}dx^{j},
   \end{array}\ee
where $h(r)$ and $f(r)$ are the non-compact extra dimension dependent metric functions with an additional constraint $h(r)f(r)=C$ with $C=1$
which is obtained from the solutions of bulk Einstein-Hillbert-Gauss-Bonnet equation provided the back reaction effect of the bulk/brane fields have been taken care of
and the Gauss-Bonnet coupling $\alpha_{(5)}<<1$. 
 Moreover in the above metric ansatz
 $\gamma_{ij}$ is the unit metric. In the field equation which follow, k denotes the curvature of ${\bf {\cal K}^{3}}$ and can take the values 1 (positive
curvature), 0 (zero curvature), and -1 (negative curvature).

\section{\bf Metric function and its asymptotic behaviour}

The brane action gives singular contribution which is addressed  by 
Israel junction conditions.
Now varying the action stated in equation(\ref{action}) and neglecting the back reaction of all the other brane fields except gravity,
the five dimensional Bulk Einstein's equation turns out to be
\be\begin{array}{lllll}\label{eneqn}
    \displaystyle G^{(5)}_{AB}+\frac{\alpha_{(5)}}{M^{2}_{(5)}}H^{(5)}_{AB}
=-\left[\frac{\Lambda_{(5)}}{M^{3}_{(5)}}+\frac{1}{2}{\cal H}_{CDE}{\cal H}^{CDE}\right]g^{(5)}_{AB}+3{\cal H}_{ACE}{\cal H}_{B}^{CE}
   \end{array}\ee
where the five dimensional Einstein's tensor and the Gauss-Bonnet tensor is given by

\be\begin{array}{llll}\label{et}
    G^{(5)}_{AB}=\left[R^{(5)}_{AB}-\frac{1}{2}g^{(5)}_{AB}R_{(5)}\right],
   \end{array}\ee

\be\begin{array}{llll}\label{gbp}
  H^{(5)}_{AB}=2R^{(5)}_{ACDE}R_{B}^{CDE(5)}-4R_{ACBD}^{(5)}R^{CD(5)}
-4R_{AC}^{(5)}R_{B}^{C(5)}+2R^{(5)}R_{AB}^{(5)}\\ ~~~~~~~~~~~~~~~~~~~~~~~~~~~~~~~~~~~~~~~~~~~-\frac{1}{2}g^{(5)}_{AB}
\left(R^{ABCD(5)}R^{(5)}_{ABCD}-4R^{AB(5)}R^{(5)}_{AB}+R^{2}_{(5)}\right).
   \end{array}\ee
To proceed further we use the fact that the rank 3 antisymmetric Kalb Ramond field strength tensor can 
be expressed in terms of the corresponding gauge potential in string theory appearing from the closed string modes as \cite{risi1,risi2}
\be\begin{array}{llll}\label{xui}
    \displaystyle {\cal H}^{ABC}=\epsilon^{ABCDE}\nabla_{D}{\cal A}_{E}.
   \end{array}\ee
Using this ansatz we have
\be\begin{array}{llll}\label{xui1}
    \displaystyle \nabla_{C}{\cal H}^{CAB}=\frac{1}{2}\epsilon^{ABCDE}\nabla_{[C}\nabla_{D]}{\cal A}_{E}=\frac{1}{4}\epsilon^{ABCDE}R_{CDE}^{M(5)}{\cal A}_{M}=0
   \end{array}\ee
which follows from the Bianchi identity $R_{[ABC]}^{D(5)}=0$. Additionally we have
\be\begin{array}{llll}\label{kjo1}
    \displaystyle {\cal H}_{ABC}{\cal H}^{ABC}=3!2!\delta_{[P}^{D}\delta_{Q]}^{E}\nabla_{D}{\cal A}_{E}\nabla^{P}{\cal A}^{Q}=12{\cal B}_{MN}{\cal B}^{MN},
   \end{array}\ee
\be\begin{array}{llll}\label{kjo2}
    \displaystyle {\cal H}_{AMN}{\cal H}_{B}^{MN}=1!2!g^{(5)}_{BQ}\delta_{[A}^{Q}\delta_{D}^{R}\delta_{E]}^{S}\nabla_{R}{\cal A}_{S}\nabla^{D}{\cal A}^{E}
=\left({\cal B}_{MN}{\cal B}^{MN}g^{(5)}_{AB}-2{\cal B}_{AC}{\cal B}_{B}^{C}\right)
   \end{array}\ee
with rank 2 antisymmetric Kalb Ramond tensor potential ${\cal B}_{MN}=-{\cal B}_{NM}$, usually called ``Neveu-Schwarz Neveu-Schwarz'' (NS-NS) two-
form. For historical reasons the field ${\cal B}$ is also called ``torsion'' since, to lowest order, it can be identified with the
antisymmetric part of the affine connection, in the context of a non-Riemannian geometric structure. An alternative,
often used, name is ``Kalb-Ramond axion'', in reference to the pseudo-scalar axionic field related to the Kalb-Ramond
antisymmetric tensor field via space-time ``duality'' transformation \cite{sayan1,ssg1,ssg2,ssg3,ssg4,ssg5}. 

Using equation(\ref{kjo1}) and equation(\ref{kjo2}) in equation(\ref{eneqn}) the Einstein's equation in terms of the Kalb Ramond two form turns out to be
\be\begin{array}{lllll}\label{eneqnz1}
    \displaystyle G^{(5)}_{AB}+\frac{\alpha_{(5)}}{M^{2}_{(5)}}H^{(5)}_{AB}
=-\left[\frac{\Lambda_{(5)}}{M^{3}_{(5)}}+3{\cal B}_{MN}{\cal B}^{MN}\right]g^{(5)}_{AB}-6{\cal B}_{AC}{\cal B}_{B}^{C}.
   \end{array}\ee
Now we assume that the Kalb Ramond gauge field is purely electric i.e. ${\cal A}_{M}=(\Phi_{KR}(r),0,0,0,0)$.
From the equation(\ref{eneqnz1}) $(A=i,B=i)$ component of Einstein's equation can be written as:
\be\begin{array}{llll}\label{gbf}
  \displaystyle  \left\{rh(r)\left[\frac{4\alpha_{(5)}}{M^{2}_{(5)}}h(r)-r^{2}-4\alpha_{(5)}k\right]-4r^{3}\right\}\left(\frac{dh(r)}{dr}\right)
+\frac{2\alpha_{(5)}r^{2}}{M^{2}_{(5)}}\left(\frac{dh(r)}{dr}\right)^{2}+2kr^{2}-2r^{2}h(r)\\ \displaystyle~~~~~~~~~~~~~~~~~~~~~~~
~~~~~~~~~~~~~~~~~~~~~~~~~~~~~~~~~~~~~~~~~~~~~~~~~~~~~~~~~~~~~~~~~~~~~~~~-\frac{2\Lambda_{(5)}r^{4}}{M^{3}_{(5)}}+12r^{4}
\left(\frac{d\Phi_{KR}(r)}{dr}\right)^{2}=0
   \end{array}\ee
and from $(A=t,B=t)$ or $(A=r,B=r)$ component we get
\be\begin{array}{llll}\label{ffge}
   \displaystyle  3r\left[\frac{4\alpha_{(5)}}{M^{2}_{(5)}}h(r)-r^{2}-4\alpha_{(5)}k\right]\left(\frac{dh(r)}{dr}\right)
-6r^{2}h(r)+6kr^{2}-\frac{2\Lambda_{(5)}r^{4}}{M^{3}_{(5)}}+24r^{4}
\left(\frac{d\Phi_{KR}(r)}{dr}\right)^{2}=0.
 \end{array}\ee

\begin{figure}[htb]
{\centerline{\includegraphics[width=13cm, height=9cm] {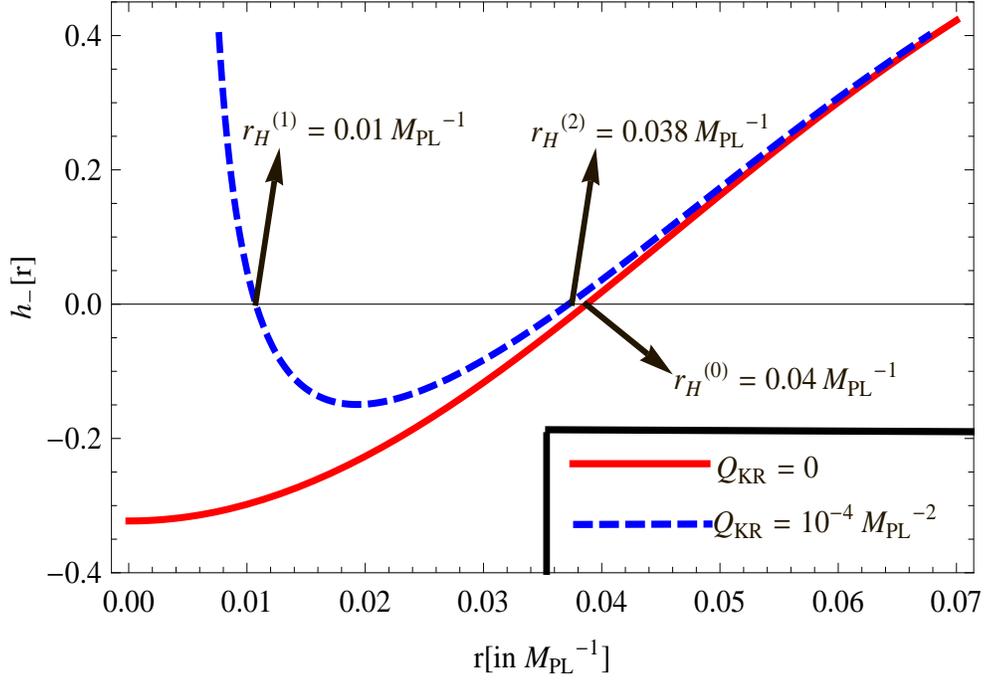}}}
\caption{Variation
 of the metric function $h_{-}(r)$
with r  
for Gauss-Bonnet coupling $\alpha_{(5)}=0.001>0$ and ADM mass parameter $\mu>0$  
in presence of Kalb Ramond field. Here we use $\mu=0.028,\Lambda_{(5)}=-1$ in the Planckian unit.} \label{fig1}
\end{figure}

Now using the additional constraint stated in equation(\ref{xui1}) the Kalb Ramond electic potential turns out to be $\Phi_{KR}(r)=\frac{Q_{KR}}{r^{2}}$,
 where $Q_{KR}$ be the Kalb Ramond charge. Substituting this result in equation(\ref{gbf}) and equation(\ref{ffge}),
 the metric function can be obtained as:
\be\begin{array}{lll}\label{hor1}
    \displaystyle h(r)=k+\frac{r^{2}M^{2}_{(5)}}{4\alpha_{(5)}}\left[1\mp\sqrt{1+\frac{\mu\alpha_{(5)}}{r^{4}}+\frac{\alpha_{(5)}}{M^{5}_{(5)}}
\left(\frac{4\Lambda_{(5)}}{3}-\frac{128Q^{2}_{KR}M^{3}_{(5)}}{r^{6}}\right)}~\right]:=h_{\mp}(r)
   \end{array}\ee
where $\mu$ is a constant which can be expressed in terms of the global mass parameter $M_{KR}$ as $\mu=\frac{16M_{KR}}{3V^{(k)}_{3}M^{5}_{(5)}}$, where $V^{(k)}_{3}$
is the is a unit volume of 
 ${\cal K}^{3}$ if it is compact. For spherical unit volume in ${\cal K}^{3}$ we have $V^{(k)}_{3}=\frac{2\pi^{\frac{3}{2}}}{\Gamma\left(\frac{3}{2}\right)}$.
Henceforth we are interested in the flat space-time ($k=1$) since only in this situation $M$ can be interpreted as the ADM mass of the black hole.
In this article we restrict the signature of the five dimensional bulk cosmological constant to be $\Lambda_{(5)}<0$ because we want to explore the AdS/CFT correspondence
from the holographic four dimensional CFT dual of the five dimensional Gauss-Bonnet gravity. So in subsequent numerical estimation we only consider $\Lambda_{(5)}<0$.
The asymptotic behaviour of the the two solutions of the metric functions for $k=1$ are given below:
\be\begin{array}{llll}\label{ho1}
    \displaystyle \lim_{\alpha_{(5)}\rightarrow 0}h_{-}(r)=\left[1-\frac{\mu M^{2}_{(5)}}{8r^{2}}
-\frac{\Lambda_{(5)}r^{2}}{6M^{3}_{(5)}}+\frac{16Q^{2}_{KR}}{r^{4}}\right]=:h_{\bf GR}(r),
   \end{array}\ee
\be\begin{array}{llll}\label{ho2}
    \displaystyle \lim_{\alpha_{(5)}\rightarrow 0}h_{+}(r)=\lim_{\alpha_{(5)}\rightarrow 0}\left[1+
\frac{\mu M^{2}_{(5)}}{8r^{2}}+
\left(\frac{\Lambda_{(5)}}{6M^{3}_{(5)}}+\frac{M^{2}_{(5)}}{2\alpha_{(5)}}\right)r^{2}-\frac{16Q^{2}_{KR}}{r^{4}}~\right]=:h_{\bf NGR}(r).
   \end{array}\ee
where ${\bf GR}$ and ${\bf NGR}$ stands for General Relativistic branch and Non-General Relativistic branch respectively.
The maximally symmetric Kalb Ramond black hole in the Non-General Relativistic branch is unstable compared to the General Relativistic branch. 
Most importantly 
in the $\alpha_{(5)}\rightarrow 0$ limit the ${\bf GR}$ branch asymptotically reaches the Schwarzchild solution 
in presence of electrical charge (Reisner-N$\ddot{o}$rdstorm type). Now from detailed numerical analysis  
we see that the +ve branch of the metric function $h_{+}(r)$ does not incorporate any horizon for both the signatures of five dimensional 
cosmological constant $\Lambda_{(5)}$, ADM mass parameter $\mu>0$ , $Q_{KR}\neq 0$ or $Q_{KR}=0$. This corresponds to the naked singular solution
 which violates the cosmic censorship. But from the -ve branch solution of the metric function
$h_{-}(r)$ we calculate the killing horizon $r_{H}$. To avoid the naked singularity in the present work we will only focus on the -ve branch solution 
of the metric function $h_{-}(r)$.
 In figure(\ref{fig1}) we have clearly shown the 
behaviour of the metric function $h_{-}(r)$ with respect to the five dimensional coordinate $r$ for -ve signature of five dimensional 
cosmological constant $\Lambda_{(5)}$. From figure(\ref{fig1}) with $\Lambda_{(5)}<0$ the numerical roots for the 
killing horizon are given by $r^{(0)}_{H}=0.04~M^{-1}_{PL}$ (for $Q_{KR}=0$), $r^{(1)}_{H}=0.01~M^{-1}_{PL}$ and
 $r^{(2)}_{H}=0.038~M^{-1}_{PL}$ (for $Q_{KR}=10^{-4}~M^{-2}_{PL}$). 

\section{\bf Branch singularity and killing horizon}

First of all it is important to mention here that for the above mentioned space-time there are two classes of curvature singularities for
$\mu\neq 0$, $\alpha_{(5)}\neq 0$ and $Q_{KR}\neq 0$. One of them is the well known ``central singularity'' at $r=0$ and the 
other is the ``branch singularity'' at $r=r_{b}(>0)$, where the term
inside the square-root in the metric function stated in equation(\ref{hor1}) for $k=1$ vanishes and the the corresponding ``branch singularity''
satisfies the following algebraic equation:
\be\begin{array}{llll}\label{aeq}
    \displaystyle \Delta^{3}+\left[\frac{\mu_{b}\alpha_{(5)}}{\left(1+\frac{4\alpha_{(5)}\Lambda_{(5)}}{3M^{5}_{(5)}}\right)}\right]\Delta
-\left[\frac{\frac{128\alpha_{(5)}Q^{2}_{KR}}{M^{2}_{(5)}}}{\left(1+\frac{4\alpha_{(5)}\Lambda_{(5)}}{3M^{5}_{(5)}}\right)}\right]=0
   \end{array}\ee
where we introduce $\Delta:=r^{2}_{b}$ and $\mu_{b}:=\mu(r=r_{b})$. The analytical solutions of equation(\ref{aeq})
 for different physical situations where no naked singularity appears are discussed below:\\
\underline{\bf Case I}:-For $4\left[\frac{\mu_{b}\alpha_{(5)}}{\left(1+\frac{4\alpha_{(5)}\Lambda_{(5)}}{3M^{5}_{(5)}}\right)}\right]^{3}
+27\left[-\frac{\frac{128\alpha_{(5)}Q^{2}_{KR}}{M^{2}_{(5)}}}{\left(1+\frac{4\alpha_{(5)}\Lambda_{(5)}}{3M^{5}_{(5)}}\right)}\right]^{2}>0$
and $\left[\frac{\mu_{b}\alpha_{(5)}}{\left(1+\frac{4\alpha_{(5)}\Lambda_{(5)}}{3M^{5}_{(5)}}\right)}\right]<0$\\
\be\begin{array}{llll}\label{cs1}
    \displaystyle \Delta=\frac{\left|-\frac{128\alpha_{(5)}Q^{2}_{KR}}{M^{2}_{(5)}}\right|}{\left(\frac{64\alpha_{(5)}Q^{2}_{KR}}{M^{2}_{(5)}}\right)}
\sqrt{-\frac{\mu_{b}\alpha_{(5)}}{3\left(1+\frac{4\alpha_{(5)}\Lambda_{(5)}}{3M^{5}_{(5)}}\right)}}\cosh\left[\frac{1}{3}\cosh^{-1}\left(-\frac{3}{2}
\frac{\left|-\frac{\frac{128\alpha_{(5)}Q^{2}_{KR}}{M^{2}_{(5)}}}{\left(1+\frac{4\alpha_{(5)}\Lambda_{(5)}}{3M^{5}_{(5)}}\right)}\right|}{\frac{\mu_{b}\alpha_{(5)}}{\left(1+\frac{4\alpha_{(5)}\Lambda_{(5)}}{3M^{5}_{(5)}}\right)}}
\sqrt{-\frac{3\left(1+\frac{4\alpha_{(5)}\Lambda_{(5)}}{3M^{5}_{(5)}}\right)}{\mu_{b}\alpha_{(5)}}}\right)\right]
   \end{array}\ee
\underline{\bf Case II}:-For $4\left[\frac{\mu_{b}\alpha_{(5)}}{\left(1+\frac{4\alpha_{(5)}\Lambda_{(5)}}{3M^{5}_{(5)}}\right)}\right]^{3}
+27\left[-\frac{\frac{128\alpha_{(5)}Q^{2}_{KR}}{M^{2}_{(5)}}}{\left(1+\frac{4\alpha_{(5)}\Lambda_{(5)}}{3M^{5}_{(5)}}\right)}\right]^{2}>0$
and $\left[-\frac{\frac{128\alpha_{(5)}Q^{2}_{KR}}{M^{2}_{(5)}}}{\left(1+\frac{4\alpha_{(5)}\Lambda_{(5)}}{3M^{5}_{(5)}}\right)}\right]<0$\\
\be\begin{array}{llll}\label{csa}
    \displaystyle \Delta=2
\sqrt{-\frac{\mu_{b}\alpha_{(5)}}{3\left(1+\frac{4\alpha_{(5)}\Lambda_{(5)}}{3M^{5}_{(5)}}\right)}}\cosh\left[\frac{1}{3}\cosh^{-1}\left(-\frac{3}{2}
\frac{\left|-\frac{\frac{128\alpha_{(5)}Q^{2}_{KR}}{M^{2}_{(5)}}}{\left(1+\frac{4\alpha_{(5)}\Lambda_{(5)}}{3M^{5}_{(5)}}\right)}\right|}{\frac{\mu_{b}\alpha_{(5)}}{\left(1+\frac{4\alpha_{(5)}\Lambda_{(5)}}{3M^{5}_{(5)}}\right)}}
\sqrt{-\frac{3\left(1+\frac{4\alpha_{(5)}\Lambda_{(5)}}{3M^{5}_{(5)}}\right)}{\mu_{b}\alpha_{(5)}}}\right)\right]
   \end{array}\ee\\
\underline{\bf Case III}:-For $0<\left(-\frac{3}{2}
\frac{\left|-\frac{\frac{128\alpha_{(5)}Q^{2}_{KR}}{M^{2}_{(5)}}}{\left(1+\frac{4\alpha_{(5)}\Lambda_{(5)}}{3M^{5}_{(5)}}\right)}\right|}{\frac{\mu_{b}\alpha_{(5)}}{\left(1+\frac{4\alpha_{(5)}\Lambda_{(5)}}{3M^{5}_{(5)}}\right)}}
\sqrt{-\frac{3\left(1+\frac{4\alpha_{(5)}\Lambda_{(5)}}{3M^{5}_{(5)}}\right)}{\mu_{b}\alpha_{(5)}}}\right)<1$ and $\left[\frac{\mu_{b}\alpha_{(5)}}{\left(1+\frac{4\alpha_{(5)}
\Lambda_{(5)}}{3M^{5}_{(5)}}\right)}\right]<0$
\be\begin{array}{llll}\label{bn1}
    \displaystyle \Delta_{(1)}=2
\sqrt{-\frac{\mu_{b}\alpha_{(5)}}{3\left(1+\frac{4\alpha_{(5)}\Lambda_{(5)}}{3M^{5}_{(5)}}\right)}}\cos\left[\frac{1}{3}\cos^{-1}\left(\frac{3}{2}
\frac{\left|-\frac{\frac{128\alpha_{(5)}Q^{2}_{KR}}{M^{2}_{(5)}}}{\left(1+\frac{4\alpha_{(5)}\Lambda_{(5)}}{3M^{5}_{(5)}}\right)}\right|}{\frac{\mu_{b}\alpha_{(5)}}{\left(1+\frac{4\alpha_{(5)}\Lambda_{(5)}}{3M^{5}_{(5)}}\right)}}
\sqrt{-\frac{3\left(1+\frac{4\alpha_{(5)}\Lambda_{(5)}}{3M^{5}_{(5)}}\right)}{\mu_{b}\alpha_{(5)}}}\right)\right]
   \end{array}\ee
\underline{\bf Case IV}:-For $1<\left(-\frac{3}{2}
\frac{\left|-\frac{\frac{128\alpha_{(5)}Q^{2}_{KR}}{M^{2}_{(5)}}}{\left(1+\frac{4\alpha_{(5)}\Lambda_{(5)}}{3M^{5}_{(5)}}\right)}\right|}{\frac{\mu_{b}\alpha_{(5)}}{\left(1+\frac{4\alpha_{(5)}\Lambda_{(5)}}{3M^{5}_{(5)}}\right)}}
\sqrt{-\frac{3\left(1+\frac{4\alpha_{(5)}\Lambda_{(5)}}{3M^{5}_{(5)}}\right)}{\mu_{b}\alpha_{(5)}}}\right)<0$ 
\be\begin{array}{llll}\label{bn2}
    \displaystyle \Delta=2
\sqrt{\frac{\mu_{b}\alpha_{(5)}}{3\left(1+\frac{4\alpha_{(5)}\Lambda_{(5)}}{3M^{5}_{(5)}}\right)}}\cos\left[\frac{1}{3}\cos^{-1}\left(-\frac{3}{2}
\frac{\left|-\frac{\frac{128\alpha_{(5)}Q^{2}_{KR}}{M^{2}_{(5)}}}{\left(1+\frac{4\alpha_{(5)}\Lambda_{(5)}}{3M^{5}_{(5)}}\right)}\right|}{\frac{\mu_{b}\alpha_{(5)}}{\left(1+\frac{4\alpha_{(5)}\Lambda_{(5)}}{3M^{5}_{(5)}}\right)}}
\sqrt{-\frac{3\left(1+\frac{4\alpha_{(5)}\Lambda_{(5)}}{3M^{5}_{(5)}}\right)}{\mu_{b}\alpha_{(5)}}}\right)\right]
   \end{array}\ee\\
\underline{\bf Case V}:-For $\left(-\frac{3}{2}
\frac{\left|-\frac{\frac{128\alpha_{(5)}Q^{2}_{KR}}{M^{2}_{(5)}}}{\left(1+\frac{4\alpha_{(5)}\Lambda_{(5)}}{3M^{5}_{(5)}}\right)}\right|}{\frac{\mu_{b}\alpha_{(5)}}{\left(1+\frac{4\alpha_{(5)}\Lambda_{(5)}}{3M^{5}_{(5)}}\right)}}
\sqrt{-\frac{3\left(1+\frac{4\alpha_{(5)}\Lambda_{(5)}}{3M^{5}_{(5)}}\right)}{\mu_{b}\alpha_{(5)}}}\right)<1$ 
\be\begin{array}{llll}\label{bn3}
    \displaystyle \Delta=2
\sqrt{\frac{\mu_{b}\alpha_{(5)}}{3\left(1+\frac{4\alpha_{(5)}\Lambda_{(5)}}{3M^{5}_{(5)}}\right)}}\cosh\left[\frac{1}{3}\cosh^{-1}\left(-\frac{3}{2}
\frac{\left|-\frac{\frac{128\alpha_{(5)}Q^{2}_{KR}}{M^{2}_{(5)}}}{\left(1+\frac{4\alpha_{(5)}\Lambda_{(5)}}{3M^{5}_{(5)}}\right)}\right|}{\frac{\mu_{b}\alpha_{(5)}}{\left(1+\frac{4\alpha_{(5)}\Lambda_{(5)}}{3M^{5}_{(5)}}\right)}}
\sqrt{-\frac{3\left(1+\frac{4\alpha_{(5)}\Lambda_{(5)}}{3M^{5}_{(5)}}\right)}{\mu_{b}\alpha_{(5)}}}\right)\right]
   \end{array}\ee
 Now combining all of these allowed solution for the ``branch singularity'' in general we can write:
\be\begin{array}{llll}\label{bn4}
    \displaystyle r_{b}=\sqrt[4]{\frac{\mu_{b}\alpha_{(5)}}{3\left(1+\frac{4\alpha_{(5)}\Lambda_{(5)}}{3M^{5}_{(5)}}\right)}}
\sqrt{{\cal C}_{\frac{1}{3}}\left[-
\frac{3\left|-\frac{\frac{128\alpha_{(5)}Q^{2}_{KR}}{M^{2}_{(5)}}}{\left(1+\frac{4\alpha_{(5)}\Lambda_{(5)}}{3M^{5}_{(5)}}\right)}\right|}{\frac{\mu_{b}\alpha_{(5)}}{\left(1+\frac{4\alpha_{(5)}\Lambda_{(5)}}{3M^{5}_{(5)}}\right)}}
\sqrt{-\frac{3\left(1+\frac{4\alpha_{(5)}\Lambda_{(5)}}{3M^{5}_{(5)}}\right)}{\mu_{b}\alpha_{(5)}}}\right]}
   \end{array}\ee
where ${\cal C}_{\frac{1}{3}}(t)$ be the {\it Chebyshev polynomial} with argument $t$ \cite{risi1,risi2}. Let us now 
concentrate on the ``killing horizon'' which has important physical significances in the context of phase transition and 
critical phenomena in the black hole thermodynamics \cite{diba1,diba2,diba3,ssg6,ssg7,ssg8,ssg9,ssg10,ssg11}. By setting $h(r_{H})=0$ in equation(\ref{hor1}) in $k=1$ we get:
\be\begin{array}{lllll}\label{op1}
    \displaystyle \Theta^{3}-\Omega _1\Theta^{2}
+\Omega _2\Theta -\Omega _3=0
   \end{array}\ee
where we introduce $\Theta:=r^{2}_{H}$ and $\mu_{H}:=\mu(r=r_{H})$. In this context we define 
\be\begin{array}{llll}\label{def1}
    \displaystyle \Omega _1=\frac{6M^{3}_{(5)}}{\Lambda_{(5)}},~~~~~~~\Omega _2=\left(\frac{3M^{5}_{(5)}}{4\Lambda_{(5)}}\left[\mu_{H}-\frac{16\alpha_{(5)}}{M^{4}_{(5)}}\right]\right),
~~~~~~~\Omega _3=\frac{96Q^{2}_{KR}M^{3}_{(5)}}{\Lambda_{(5)}}.
   \end{array}\ee
\begin{figure}[htb]
{\centerline{\includegraphics[width=12cm, height=8cm] {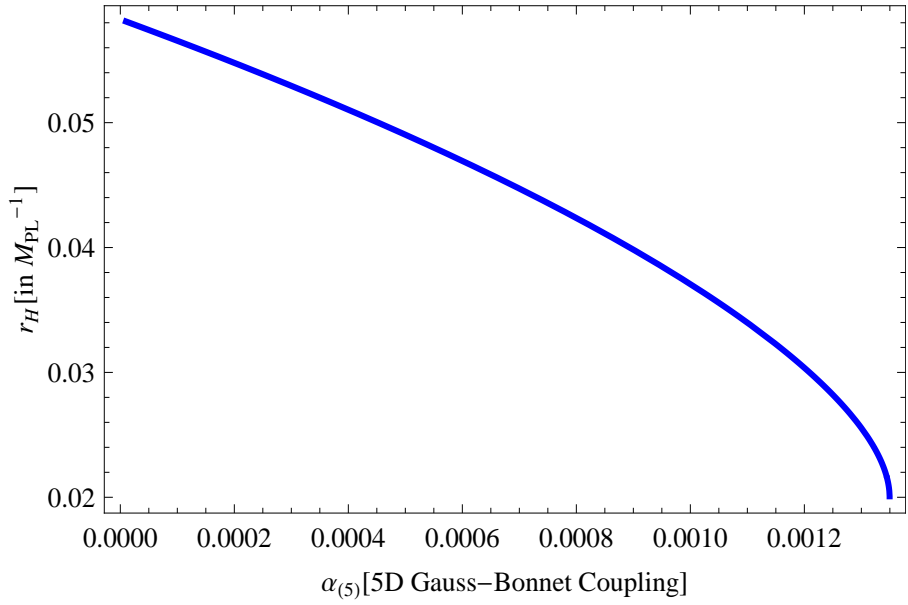}}}
\caption{Variation
 of the killing horizon $r_{H}$
vs five dimensional Gauss-Bonnet coupling 
 and ADM mass parameter $\mu=0.028>0$
in presence of Kalb Ramond field with charge $Q_{KR}=10^{-4}~M^{-2}_{PL}$.} \label{fig2}
\end{figure}
\begin{figure}[htb]
{\centerline{\includegraphics[width=12cm, height=8cm] {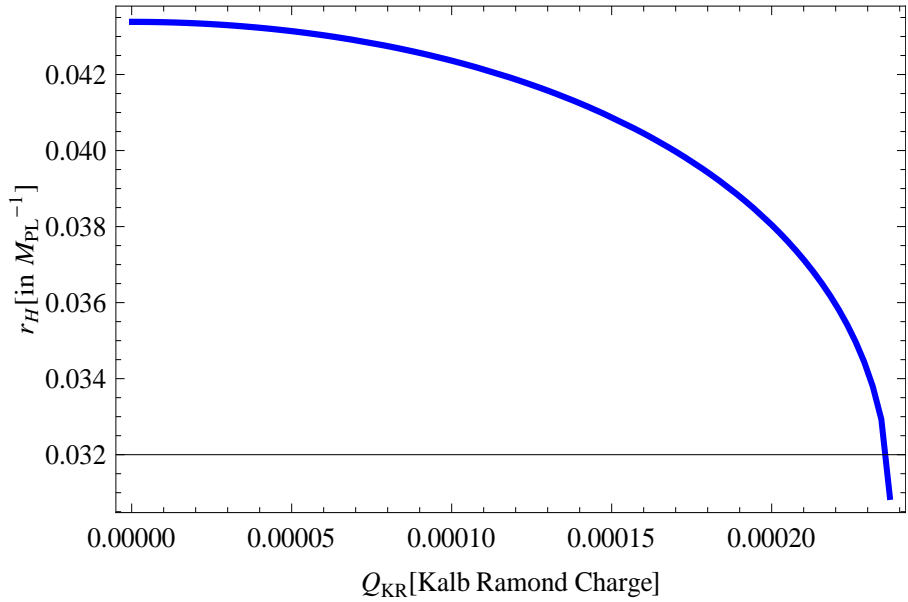}}}
\caption{Variation
 of the killing horizon $r_{H}$
with Kalb Ramond charge $Q_{KR}$ 
 and ADM mass parameter $\mu=0.028>0$  
in presence of Kalb Ramond field.} \label{fig2a}
\end{figure}

Consequently the real root of equation(\ref{op1})
  is given by:
\be\begin{array}{llll}\label{jk1}
    \displaystyle \Theta=\frac{\Omega _1}{3}-\frac{\sqrt[3]{2} \left(-\Omega _1^2+3 \Omega _2\right)}{3 \sqrt[3]{\left(2 \Omega _1^3-9 \Omega _1 \Omega _2+27 \Omega _3
+\sqrt{4 \left(-\Omega _1^2+3 \Omega _2\right){}^3+\left(2 \Omega _1^3-9 \Omega _1 \Omega _2
+27 \Omega _3\right){}^2}\right)}}\\ \displaystyle~~~~~~~~~~~~~~~~~~~~~~~~~~~~~~~~~~~~~~~~+\frac{\sqrt[3]{\left(2 \Omega _1^3-9 \Omega _1 \Omega _2
+27 \Omega _3+\sqrt{4 \left(-\Omega _1^2+3 \Omega _2\right){}^3+\left(2 \Omega _1^3-9 \Omega _1 \Omega _2+27 \Omega _3\right){}^2}\right)}}{3\sqrt[3]{ 2}}
   \end{array}\ee
which gives the physical solution for the ``killing horizon''. Using equation(\ref{jk1}) the characteristic features as well as the phase transition
phenomena of charged Kalb Ramond black hole is elaborately discussed in the next sections. Most importantly
in $\alpha_{(5)}\rightarrow 0$ asymptotic limit the expression for the ``killing horizon'' is almost same but the expression for $\Omega_{2}$ is modified.
Same situation appears for the calculation of ``branch singularity'' also. In figure(\ref{fig2}) we have shown the functional dependence of killing horizon $r_{H}$
with respect to the five dimensional Gauss-Bonnet coupling $\alpha_{(5)}$ for a fixed ADM mass parameter $\mu$ and Kalb Ramond black hole charge $Q_{KR}$. 
This clearly shows that as the Gauss-Bonnet coupling increases, the corresponding numerical value of the killing horizon decreases. We have also shown the behaviour 
of killing horizon $r_{H}$ with respect to the Kalb Ramond charge $Q_{KR}$ with fixed numerical value of $\alpha_{(5)}=0.0001$ in figure(\ref{fig2a}).
\section{\bf Thermodynamical analysis of KR-ADS Black holes }
In this section we derive different thermodynamical quantities for the charged KR-ADS black hole described in the previous section.


\subsection{\bf Hawking temperature}

\begin{figure}[htb]
{\centerline{\includegraphics[width=12cm, height=8cm] {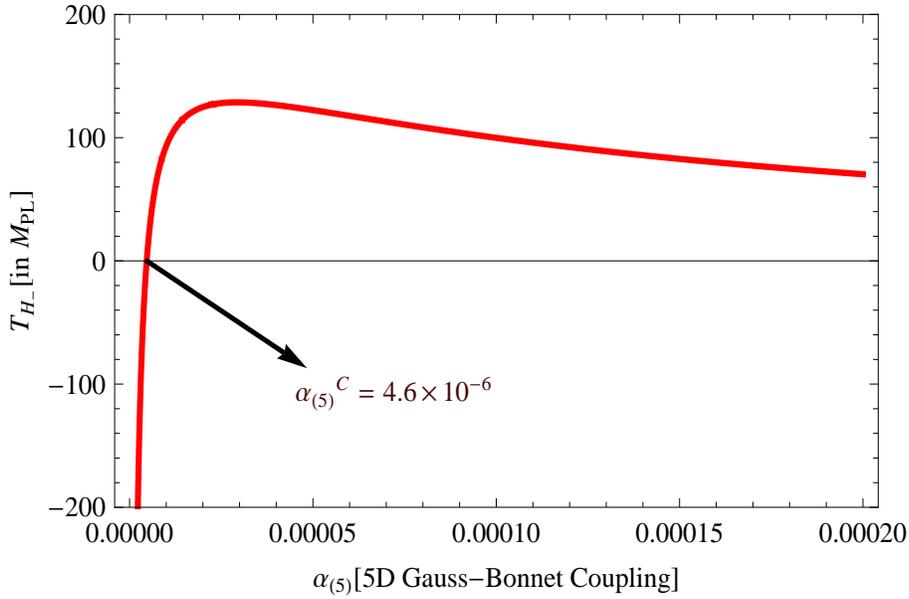}}}
\caption{Variation
 of the Hawking temperature $T_{H_{-}}$ 
with five dimensional Gauss-Bonnet coupling $\alpha_{(5)}$ for $Q_{KR}=10^{-4}$ 
in presence of Kalb Ramond antisymmetric tensor field. Here we use the ADM mass parameter $\mu=0.028$ in the Planckian unit.} \label{fig3}
\end{figure}

\begin{figure}[htb]
{\centerline{\includegraphics[width=12cm, height=8cm] {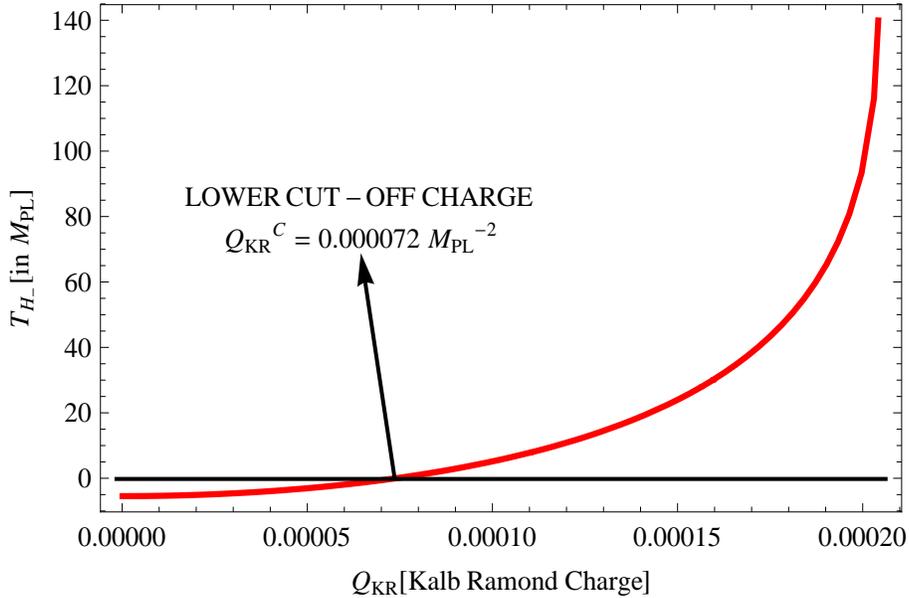}}}
\caption{Variation
 of the Hawking temperature $T_{H_{-}}$ 
with Kalb Ramond charge  $Q_{KR}$  
in presence of Kalb Ramond antisymmetric tensor field. Here we use the ADM mass parameter $\mu=0.028$ with Gauss-Bonnet coupling $\alpha_{(5)}=\alpha^{C}_{(5)}$
.} \label{fig3a}
\end{figure}

In the context of black hole thermodynamics ``Hawking temperature'' is defined as:
\be\label{iou1}
T_{-}=\frac{\kappa}{2\pi}\ee
where $\kappa$ is the ``surface gravity'' defined as:
\be\label{iou2} \kappa=\frac{1}{2}\left(\frac{dh_{-}(r)}{dr}\right)_{r=r_{H}}.\ee
Using equation(\ref{hor1}) the ``Hawking temperature'' for the charged Kalb Ramond black hole can be expressed as:
\be\begin{array}{llll}\label{iou3}
    \displaystyle T_{-}=- \frac{M^{2}_{(5)}}{16\pi\alpha_{(5)}}\left\{2r_{H}\sqrt{1+\frac{\mu\alpha_{(5)}}{r^{4}_{H}}+\frac{\alpha_{(5)}}{M^{5}_{(5)}}
\left(\frac{4\Lambda_{(5)}}{3}-\frac{128Q^{2}_{KR}M^{3}_{(5)}}{r^{6}_{H}}\right)}+\frac{\left(\frac{768Q^{2}_{KR}\alpha_{(5)}}{M^{2}_{(5)}r^{5}_{H}}
-\frac{4\mu\alpha_{(5)}}{r^{3}_{H}}\right)}{\sqrt{1+\frac{\mu\alpha_{(5)}}{r^{4}_{H}}+\frac{\alpha_{(5)}}{M^{5}_{(5)}}
\left(\frac{4\Lambda_{(5)}}{3}-\frac{128Q^{2}_{KR}M^{3}_{(5)}}{r^{6}_{H}}\right)}}\right\}
   \end{array}\ee
where the ``killing horizon'' ($r_{H}$) is calculated from equation(\ref{jk1}).

In $\alpha_{(5)}\rightarrow 0$ asymptotic limit, the expression for the ``Hawking temperature'' reduces to the 
following form:
\be\begin{array}{llll}\label{opi1}
    \displaystyle \lim_{\alpha_{(5)}\rightarrow 0}T_{-}:=T_{\bf GR}=\frac{1}{4\pi}\left[\frac{\mu M^{2}_{(5)}}{4r^{3}_{\star}}
-\frac{\Lambda_{(5)}r_{\star}}{3M^{3}_{(5)}}-\frac{64Q^{2}_{KR}}{r^{5}_{\star}}\right]
   \end{array}\ee
where $r_{\star}=r_{H}(\alpha_{(5)}\rightarrow 0)$ which is evaluated from $\lim_{\alpha_{(5)}\rightarrow 0}h_{-}(r_{\star})=0$. In figure(\ref{fig3})
we have shown the behaviour of Hawking temperature with respect to the five dimensional Gauss-Bonnet coupling $\alpha_{(5)}$. To satisfy the 
constraint appearing from third law of thermodynamics here we have to fix the lower bound on five dimensional Gauss-Bonnet coupling as explicitly
 shown in figure(\ref{fig3}).
In the present context the permissible value of the lower cut-off of $\alpha_{(5)}$ is $4.6\times 10^{-6}$ for $\Lambda_{(5)}<0$.
From the figure(\ref{fig3}) we see that as the five dimension Gauss-Bonnet coupling $\alpha_{(5)}$ changes its numerical value in the neighborhood of the lower cut-off
$\alpha^{C}_{(5)}$ from lower to higher then the corresponding Hawking temperature increases and reaches a maximum value at $\alpha_{(5)}=0.00002$. After that as 
$\alpha_{(5)}$ increases the Hawking temperature decreases. Additionally, we have also depicted the behaviour of Hawking temperature with respect to 
the Kalb Ramond charge $Q_{KR}$ for fixed value of $\alpha_{(5)}=\alpha^{C}_{((5)}$ in figure(\ref{fig3a}). The lower cut-off of Kalb Ramond charge from the figure(\ref{fig3a}) turns out to be $Q^{C}_{KR}=0.000072~M^{-2}_{PL}$.
\subsection{\bf Bekenstein Hawking entropy}

In presence of GB coupling ($\alpha_{(5)}$) the ``Bekenstein Hawking entropy'' in five dimension is defined as
 \cite{paddy1,jacobson1,simon,gen,gen1,nojiri,nojiri2,nojiri3,tim,nup,nup2,katz,olia,gen3,akbar}:
\be\label{bh1}
S_{H}=\frac{A_{BH}}{4G_{N}}\left\{1+\frac{6(2\pi^{2})^{\frac{2}{3}}\alpha_{(5)}}{A^{\frac{2}{3}}_{BH}M^{2}_{PL}}\right\}
=\frac{\pi^2}{2}\left(\frac{ Q_{KR}}{\Phi_{KR}}\right)^{\frac{3}{2}}\left\{1+\frac{6\alpha_{(5)}\Phi_{KR}}{Q_{KR}M^{2}_{PL}}\right\}\ee
where $A_{BH}$ is the area of the charged Kalb Ramond black hole defined as \cite{paddy1,jacobson1,simon,gen,gen1,nojiri,nojiri2,nojiri3,tim,nup,nup2,katz,olia,gen3,akbar}, 
$A_{BH}=\frac{2\pi^{2}}{\Gamma(2)}r^{3}_{H}=2\pi^{2}r^{3}_{H}$
and $G_{N}$ and $M_{PL}$ are the Gravitational constant which is taken to be unity in the Planckian unit respectively. Using 
equation(\ref{jk1}) the ``Bekenstein Hawking entropy'' for charged Kalb Ramond black hole can be expressed as:

\begin{figure}[htb]
{\centerline{\includegraphics[width=12cm, height=8cm] {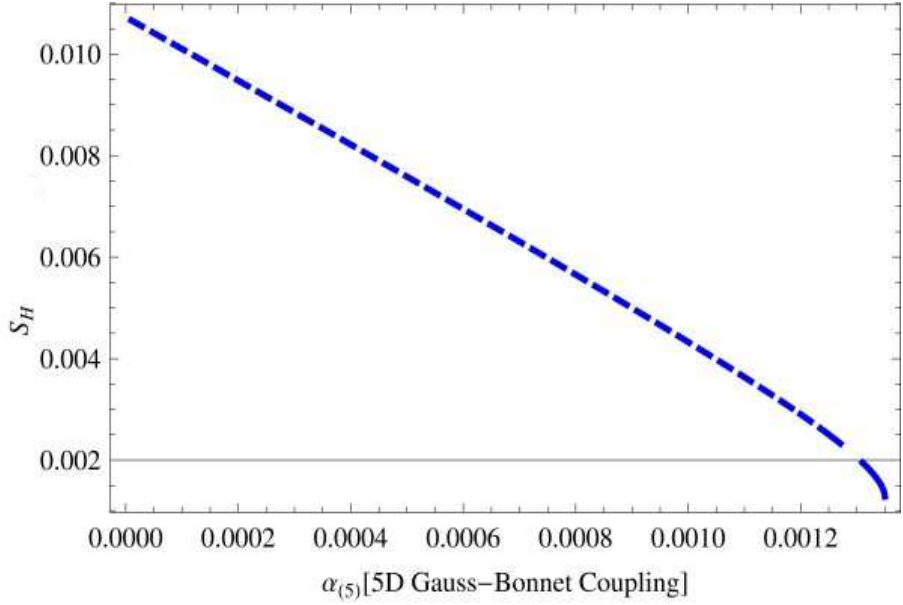}}}
\caption{Variation
 of Bekenstein Hawking entropy ($S_{H}$) 
with five dimensional Gauss-Bonnet coupling $\alpha_{(5)}$ 
in presence of charged Kalb Ramond antisymmetric tensor field with charge $Q_{KR}=10^{-4}M^{-2}_{PL}$. Here we use $\mu=0.028$ 
in the Planckian unit.} \label{fig4}
\end{figure}

\begin{figure}[htb]
{\centerline{\includegraphics[width=12cm, height=8cm] {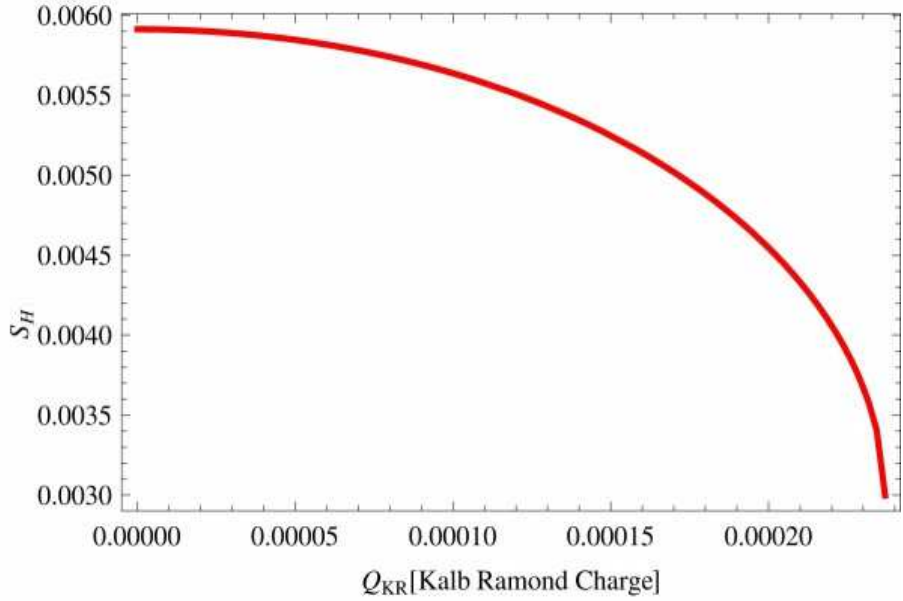}}}
\caption{Variation
 of Bekenstein Hawking entropy ($S_{H}$) 
with Kalb Ramond charge $Q_{KR}$ five dimensional Gauss-Bonnet coupling $\alpha_{(5)}=0.004$.
in the Planckian unit.} \label{fig4a}
\end{figure}

\be\begin{array}{lll}\label{kl23}
 \displaystyle  S_{H}=\frac{\pi^{2}}{2}\left\{\frac{\Omega _1}{3}-\frac{\sqrt[3]{2} \left(-\Omega _1^2+3 \Omega _2\right)}{3 \sqrt[3]{\left(2 \Omega _1^3-9 \Omega _1 \Omega _2+27 \Omega _3
+\sqrt{4 \left(-\Omega _1^2+3 \Omega _2\right){}^3+\left(2 \Omega _1^3-9 \Omega _1 \Omega _2
+27 \Omega _3\right){}^2}\right)}}\right.\\\left. \displaystyle~~~~~~~~~~~~~~~~~~~~~~~~~~~~~~~~~~~~~~~~
+\frac{\sqrt[3]{\left(2 \Omega _1^3-9 \Omega _1 \Omega _2
+27 \Omega _3+\sqrt{4 \left(-\Omega _1^2+3 \Omega _2\right){}^3+\left(2
 \Omega _1^3-9 \Omega _1 \Omega _2+27 \Omega _3\right){}^2}\right)}}{3\sqrt[3]{ 2}}\right\}^{\frac{3}{2}}\\
\displaystyle~~~~~~~~~~~~~~~~~~\times\left[1+\frac{6\alpha_{(5)}}{M^{2}_{PL}}\left\{\frac{\Omega _1}{3}-\frac{\sqrt[3]{2} \left(-\Omega _1^2+3 \Omega _2\right)}{3 \sqrt[3]{\left(2 \Omega _1^3-9 \Omega _1 \Omega _2+27 \Omega _3
+\sqrt{4 \left(-\Omega _1^2+3 \Omega _2\right){}^3+\left(2 \Omega _1^3-9 \Omega _1 \Omega _2
+27 \Omega _3\right){}^2}\right)}}\right.\right.\\ \left.\left. \displaystyle~~~~~~~~~~~~~~~~~~~~~~~~~~~~~~~~~~~~~~~~
+\frac{\sqrt[3]{\left(2 \Omega _1^3-9 \Omega _1 \Omega _2
+27 \Omega _3+\sqrt{4 \left(-\Omega _1^2+3 \Omega _2\right){}^3+\left(2
 \Omega _1^3-9 \Omega _1 \Omega _2+27 \Omega _3\right){}^2}\right)}}{3\sqrt[3]{ 2}}\right\}^{-1}\right]
   \end{array}\ee
where the constants $\Omega _1,~ \Omega _2,~\Omega _3$ are defined in equation(\ref{def1}). In the asymptotic limit $\alpha_{(5)}\rightarrow 0$
the expression for the entropy can be calculated with $\lim_{\alpha_{(5)}\rightarrow 0}\Omega _2:=\Psi=\frac{3M^{5}_{(5)}\mu_{H}}{4\Lambda_{(5)}}$.
In figure(\ref{fig4}) we have shown the behaviour of Bekenstein Hawking entropy $S_{H}$ with respect to the five dimensional Gauss-Bonnet coupling.
 Here we see that as the numerical value of $\alpha_{(5)}$ increases, the corresponding
Bekenstein Hawking entropy decreases. We have also shown the behaviour of Bekenstein Hawking entropy with respect the Kalb Ramond charge $Q_{KR}$
for a fixed value of $\alpha_{(5)}=0.004$ in figure(\ref{fig4a}).

\subsection{\bf Specific heat at constant Kalb Ramond charge}
In this context the specific heat at constant Kalb Ramond electric charge is defined as:
\be\label{sp1}
C_{Q_{KR}}=T_{-}\left(\frac{\partial S_{H}}{\partial T_{-}}\right)_{Q_{KR}}=T_{-}\frac{\left(\frac{\partial S_{H}}{\partial r_{H}}\right)_{Q_{KR}}}
{\left(\frac{\partial T_{-}}{\partial r_{H}}\right)_{Q_{KR}}}
\ee
Using equation(\ref{iou3}) and equation(\ref{kl23}) the expression for the specific heat turns out to be:
\be\begin{array}{lllll}\label{a1}
 \displaystyle C^{-}_{Q_{KR}}=\frac{{\cal B}_{1}\left(\alpha_{(5)},\Lambda_{(5)},Q_{KR},\mu,r_{H}\right)}
{{\cal B}_{2}\left(\alpha_{(5)},\Lambda_{(5)},Q_{KR},\mu,r_{H}\right)}
   \end{array}\ee
where 
\be\begin{array}{llll}\label{ra1}
    \displaystyle {\cal B}_{1}\left(\alpha_{(5)},\Lambda_{(5)},Q_{KR},\mu,r_{H}\right):=\left[\frac{3}{2}\pi^{2} r^{2}_{H}
+3\alpha_{(5)}\pi^2\right]\left\{2r_{H}\sqrt{1+\frac{\mu\alpha_{(5)}}{r^{4}_{H}}+\frac{\alpha_{(5)}}{M^{5}_{(5)}}
\left(\frac{4\Lambda_{(5)}}{3}-\frac{128Q^{2}_{KR}M^{3}_{(5)}}{r^{6}_{H}}\right)}\right.\\ \left. \displaystyle ~~~~~~~~~~~~~~~~~~~~~~~~~~~~~~~~~~~~~~~~~~~~~~~~~~~~~~~~~~~~~~~~~~~~~~~~~~~
+
\frac{\left(\frac{768Q^{2}_{KR}\alpha_{(5)}}{M^{2}_{(5)}r^{5}_{H}}-\frac{4\mu\alpha_{(5)}}{r^{3}_{H}}\right)}{\sqrt{1+\frac{\mu\alpha_{(5)}}{r^{4}_{H}}+\frac{\alpha_{(5)}}{M^{5}_{(5)}}
\left(\frac{4\Lambda_{(5)}}{3}-\frac{128Q^{2}_{KR}M^{3}_{(5)}}{r^{6}_{H}}\right)}}\right\}
   \end{array}\ee
\be\begin{array}{llll}\label{ra2}
    \displaystyle {\cal B}_{2}\left(\alpha_{(5)},\Lambda_{(5)},Q_{KR},\mu,r_{H}\right):=\left\{2\sqrt{1+\frac{\mu\alpha_{(5)}}{r^{4}_{H}}+\frac{\alpha_{(5)}}{M^{5}_{(5)}}
\left(\frac{4\Lambda_{(5)}}{3}-\frac{128Q^{2}_{KR}M^{3}_{(5)}}{r^{6}_{H}}\right)}
\right.\\ \left. \displaystyle~~~~~~~~~~~~~~~~~~~~~~~~~~
-\frac{\left(\frac{3072Q^{2}_{KR}\alpha_{(5)}}{M^{2}_{(5)}r^{6}_{H}}-\frac{8\mu\alpha_{(5)}}{r^{4}_{H}}\right)}
{\sqrt{1+\frac{\mu\alpha_{(5)}}{r^{4}_{H}}+\frac{\alpha_{(5)}}{M^{5}_{(5)}}
\left(\frac{4\Lambda_{(5)}}{3}-\frac{128Q^{2}_{KR}M^{3}_{(5)}}{r^{6}_{H}}\right)}}-
\frac{\left(\frac{768Q^{2}_{KR}\alpha_{(5)}}{M^{2}_{(5)}r^{5}_{H}}-\frac{4\mu\alpha_{(5)}}{r^{3}_{H}}\right)^{2}
}
{2r^{2}_{H}\left[1+\frac{\mu\alpha_{(5)}}{r^{4}_{H}}+\frac{\alpha_{(5)}}{M^{5}_{(5)}}
\left(\frac{4\Lambda_{(5)}}{3}-\frac{128Q^{2}_{KR}M^{3}_{(5)}}{r^{6}_{H}}\right)\right]^{\frac{3}{2}}}\right\}
   \end{array}\ee
In the asymptotic limit $\alpha_{(5)}\rightarrow 0$, equation(\ref{a1}) reduces to the following expressions:
\be\begin{array}{llll}\label{cv1}
    \displaystyle \lim_{\alpha_{(5)}\rightarrow 0}C^{-}_{Q_{KR}}=C^{\bf GR}_{Q_{KR}}=\frac{\left[\frac{3}{2}\pi^{2} r^{2}_{\star}
+3\alpha_{(5)}\pi^2\right]
\left[\frac{\mu M^{2}_{(5)}}{4r^{3}_{\star}}
-\frac{\Lambda_{(5)}r_{\star}}{3M^{3}_{(5)}}-\frac{64Q^{2}_{KR}}{r^{5}_{\star}}\right]}{\left[\frac{320Q^{2}_{KR}}{r^{6}_{\star}}-\frac{3\mu M^{2}_{(5)}}{4r^{4}_{\star}}
-\frac{\Lambda_{(5)}}{3M^{3}_{(5)}}\right]}
   \end{array}\ee


\begin{figure}[htb]
{\centerline{\includegraphics[width=12cm, height=8cm] {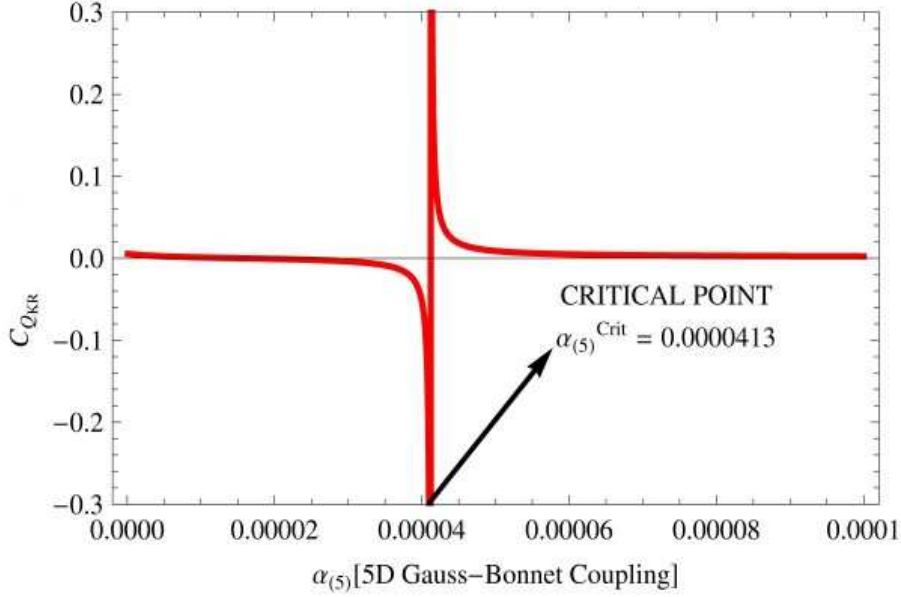}}}
\caption{Variation
 of specific heat $C^{-}_{Q_{KR}}(:=C_{Q_{KR}})$ 
with five dimensional Gauss-Bonnet coupling $\alpha_{(5)}$ Kalb Ramond charge $Q_{KR}=10^{-4}$
 Here we use ADM mass parameter $\mu=0.028$ in the Planckian unit.} \label{fig5}
\end{figure}

\begin{figure}[htb]
{\centerline{\includegraphics[width=12cm, height=8cm] {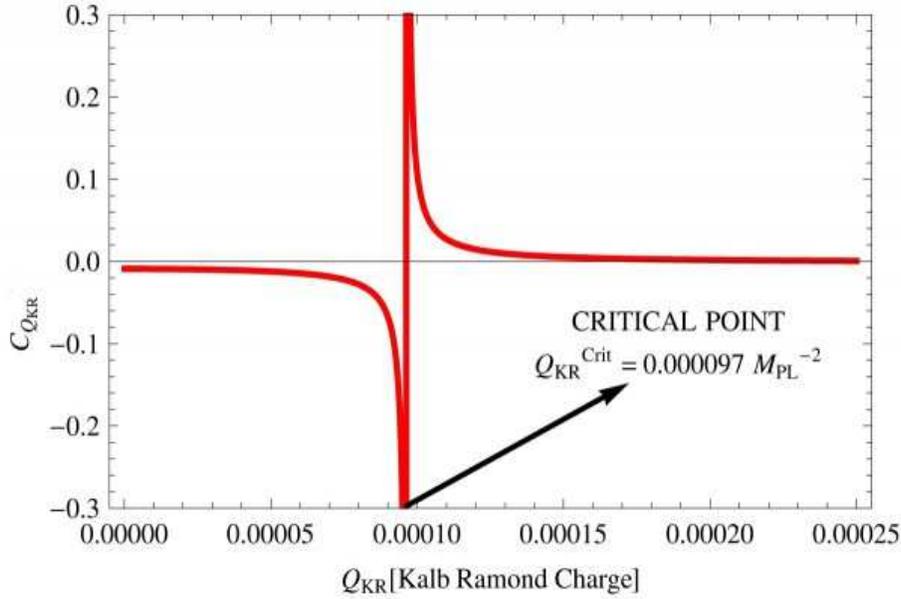}}}
\caption{Variation
 of specific heat $C^{-}_{Q_{KR}}(:=C_{Q_{KR}})$ 
with Kalb Ramond charge $Q_{KR}$ five dimensional Gauss-Bonnet coupling $\alpha_{(5)}=\alpha^{Crit}_{(5)}$.
 Here we use ADM mass parameter $\mu=0.028$ in the Planckian unit.} \label{fig5a}
\end{figure}

In $Q_{KR}\rightarrow 0$ limit the expression for the specific heat turns out to be
\be\begin{array}{llll}\label{g1}
    \displaystyle \lim_{Q_{KR}\rightarrow 0}C^{-}_{Q_{KR}}
=\frac{\left[\frac{3}{2}\pi^{2} r^{2}_{\star}
+3\alpha_{(5)}\pi^2\right]\left\{2r_{H}\sqrt{1+\frac{\mu\alpha_{(5)}}{r^{4}_{H}}+\frac{4\Lambda_{(5)}\alpha_{(5)}}{3M^{5}_{(5)}}
}-
\frac{\left(\frac{4\mu\alpha_{(5)}}{r^{3}_{H}}\right)
}{\sqrt{1+\frac{\mu\alpha_{(5)}}{r^{4}_{H}}+\frac{4\Lambda_{(5)}\alpha_{(5)}}{3M^{5}_{(5)}}}}\right\}}
{\left\{2\sqrt{1+\frac{\mu\alpha_{(5)}}{r^{4}_{H}}+\frac{4\Lambda_{(5)}\alpha_{(5)}}{3M^{5}_{(5)}}}
+\frac{\left(\frac{8\mu\alpha_{(5)}}{r^{4}_{H}}\right)}
{\sqrt{1+\frac{\mu\alpha_{(5)}}{r^{4}_{H}}+\frac{4\Lambda_{(5)}\alpha_{(5)}}{3M^{5}_{(5)}}}}-
\frac{\left(\frac{4\mu\alpha_{(5)}}{r^{3}_{H}}\right)^{2}
}
{2r^{2}_{H}\left[1+\frac{\mu\alpha_{(5)}}{r^{4}_{H}}+\frac{4\Lambda_{(5)}\alpha_{(5)}}{3M^{5}_{(5)}}\right]^{\frac{3}{2}}}\right\}}
   \end{array}\ee
and for $\alpha_{(5)}\rightarrow 0$ limit we get:
\be\begin{array}{llll}\label{g2}
    \displaystyle \lim_{
Q_{KR}\rightarrow 0}C^{\bf GR}_{Q_{KR}}=-\frac{\left[\frac{3}{2}\pi^{2} r^{2}_{\star}
+3\alpha_{(5)}\pi^2\right]
\left[\frac{\mu M^{2}_{(5)}}{4r^{3}_{\star}}
-\frac{\Lambda_{(5)}r_{\star}}{3M^{3}_{(5)}}\right]}{\left[\frac{3\mu M^{2}_{(5)}}{4r^{4}_{\star}}
+\frac{\Lambda_{(5)}}{3M^{3}_{(5)}}\right]}.
   \end{array}\ee
From equation(\ref{g2}) it is evident that when $r_{\star}<\sqrt[4]{\frac{3\mu M^{5}_{(5)}}{4\Lambda_{(5)}}}$ the corresponding specific heat is negative.
In figure(\ref{fig5}) we plot the behaviour of the specific heat at constant Kalb Ramond charge with respect to the five dimensional Gauss-Bonnet coupling
$\alpha_{(5)}$ for -ve signatures of cosmological constant $\Lambda_{(5)}$. 
The non trivial feature comes from figure(\ref{fig5}) with $Q_{KR}=10^{-4}M^{2}_{PL}$ which shows discontinuity in the specific heat $C^{-}_{Q_{KR}}$ 
with respect to the 
five dimensional Gauss-Bonnet coupling $\alpha_{(5)}$. This clearly shows the existence of phase transition in the charged Kalb Ramond black hole.
The detailed study of phase transition and critical phenomena are explicitly discussed in the next section. As an example at $\alpha_{(5)}\sim\alpha^{Crit}_{(5)}
=4.13\times 10^{-5}$, figure(\ref{fig5}) shows the existence of the phase transition in our set up. Additionally we have also shown the behaviour of the specific heat $C^{-}_{Q_{KR}}$
with respect to the Kalb Ramond charge for fixed value of $\alpha_{5)}=\alpha^{Crit}_{(5)}$ in figure(\ref{fig5a}).
\subsection{\bf Isothermal Compressibility}
In the context of black hole thermodynamics the isothermal compressibility or isothermal compression coefficient is defined as:
\be\label{com1}
K^{-1}_{T_{-}}=Q_{KR}\left(\frac{\partial \Phi_{KR}}{\partial Q_{KR}}\right)_{T_{-}}=-Q_{KR}\left(\frac{\partial \Phi_{KR}}{\partial T_{-}}\right)_{Q_{KR}}
\left(\frac{\partial T_{-}}{\partial Q_{KR}}\right)_{ \Phi_{KR}}\ee
where we use the the well known thermodynamical identity given by
\be\label{com2}
\left(\frac{\partial \Phi_{KR}}{\partial Q_{KR}}\right)_{T_{-}}\left(\frac{\partial \Phi_{KR}}{\partial T_{-}}\right)_{Q_{KR}}
\left(\frac{\partial T_{-}}{\partial Q_{KR}}\right)_{ \Phi_{KR}}=-1.\ee
To find out the isothermal compressibility for charged Kalb Ramond black hole we need to express ``Hawking temperature'' stated in equation(\ref{iou3}) in
terms of Kalb Ramond electric charge ($Q_{KR}$) and the Kalb Ramond electric potential ($\Phi_{KR}$) using the relation $\Phi_{KR}(r=r_{H})=\frac{Q_{KR}}{r^{2}_{H}}$.
Consequently equation(\ref{iou3}) takes the following form:
\be\begin{array}{llll}\label{iou31}
   T_{-}=- \frac{M^{2}_{(5)}}{16\pi\alpha_{(5)}}\left\{2\sqrt{\frac{Q_{KR}}{\Phi_{KR}}}\sqrt{1+\frac{\mu\alpha_{(5)}\Phi^{2}_{KR}}{Q^{2}_{KR}}
+\frac{\alpha_{(5)}}{M^{5}_{(5)}}
\left(\frac{4\Lambda_{(5)}}{3}-\frac{128\Phi^{3}_{KR}M^{3}_{(5)}}{Q_{KR}}\right)}
+\frac{\left(\frac{768\Phi^{\frac{5}{2}}_{KR}\alpha_{(5)}}{M^{2}_{(5)}\sqrt{Q_{KR}}}-\frac{4\mu\alpha_{(5)}\Phi^{\frac{3}{2}}}{Q^{\frac{3}{2}}_{KR}}\right)}{\sqrt{1+\frac{\mu\alpha_{(5)}\Phi^{2}_{KR}}{Q^{2}_{KR}}
+\frac{\alpha_{(5)}}{M^{5}_{(5)}}
\left(\frac{4\Lambda_{(5)}}{3}-\frac{128\Phi^{3}_{KR}M^{3}_{(5)}}{Q_{KR}}\right)}}\right\}.
   \end{array}\ee

Now using equation(\ref{iou31}) in equation(\ref{com1}) we get:
\be\begin{array}{llll}\label{l1}
   \displaystyle K^{-1}_{T_{-}}=-Q_{KR}\frac{\Xi(\alpha_{(5)},Q_{KR},\Phi_{KR},\Lambda_{(5)},\mu)}{\Sigma(\alpha_{(5)},Q_{KR},\Phi_{KR},\Lambda_{(5)},\mu)}
   \end{array}\ee
where 
\be\begin{array}{llll}\label{b1}
\Xi(\alpha_{(5)},Q_{KR},\Phi_{KR},\Lambda_{(5)},\mu):=
\left\{\frac{\sqrt{1+\frac{\mu\alpha_{(5)}\Phi^{2}_{KR}}{Q^{2}_{KR}}
+\frac{\alpha_{(5)}}{M^{5}_{(5)}}
\left(\frac{4\Lambda_{(5)}}{3}-\frac{128\Phi^{3}_{KR}M^{3}_{(5)}}{Q_{KR}}\right)}}{\sqrt{\Phi_{KR}Q_{KR}}}
+\frac{\left(\frac{8\Phi^{\frac{3}{2}}_{KR}\alpha_{(5)}}{Q^{\frac{5}{2}}_{KR}}\left[\mu+\frac{32\Phi_{KR}Q_{KR}}{M^{2}_{(5)}}\right]
-\frac{384\Phi^{\frac{5}{2}}_{KR}\alpha_{(5)}}{M^{2}_{(5)}Q^{\frac{3}{2}}_{KR}}\right)}{\sqrt{1+\frac{\mu\alpha_{(5)}\Phi^{2}_{KR}}{Q^{2}_{KR}}
+\frac{\alpha_{(5)}}{M^{5}_{(5)}}
\left(\frac{4\Lambda_{(5)}}{3}-\frac{128\Phi^{3}_{KR}M^{3}_{(5)}}{Q_{KR}}\right)}}\right.\\ \left.~~~~~~~~~~~~~~~~~~~~~~~~~~~~~~~~~~
~~~~~~~~~~~~~~~~~~~~~~~~~~~~~~~~~~~~~~~~~~~~~~~~~~~~-\frac{\left(\frac{768\Phi^{\frac{9}{2}}_{KR}\alpha^{2}_{(5)}}{M^{2}_{(5)}Q^{\frac{7}{2}}_{KR}}-\frac{4\mu\alpha_{(5)}\Phi^{\frac{3}{2}}}{Q^{\frac{3}{2}}_{KR}}
\right)\left[\mu+\frac{64\Phi_{KR}Q_{KR}}{M^{2}_{(5)}}\right]
}{\left[1+\frac{\mu\alpha_{(5)}\Phi^{2}_{KR}}{Q^{2}_{KR}}
+\frac{\alpha_{(5)}}{M^{5}_{(5)}}
\left(\frac{4\Lambda_{(5)}}{3}-\frac{128\Phi^{3}_{KR}M^{3}_{(5)}}{Q_{KR}}\right)\right]^{\frac{3}{2}}}\right\},
\end{array}\ee

\be\begin{array}{llll}\label{b2}
\Sigma(\alpha_{(5)},Q_{KR},\Phi_{KR},\Lambda_{(5)},\mu):=\left\{\frac{\sqrt{1+\frac{\mu\alpha_{(5)}\Phi^{2}_{KR}}{Q^{2}_{KR}}
+\frac{\alpha_{(5)}}{M^{5}_{(5)}}
\left(\frac{4\Lambda_{(5)}}{3}-\frac{128\Phi^{3}_{KR}M^{3}_{(5)}}{Q_{KR}}\right)}}{-\frac{\Phi^{\frac{3}{2}}}{\sqrt{Q_{KR}}}}
-\frac{\left(\frac{4\Phi^{\frac{1}{2}}_{KR}\alpha_{(5)}}{Q^{\frac{3}{2}}_{KR}}\left[\mu+\frac{48\Phi_{KR}Q_{KR}}{M^{2}_{(5)}}\right]
-\frac{1920\Phi^{\frac{3}{2}}_{KR}\alpha_{(5)}}{M^{2}_{(5)}Q^{\frac{3}{2}}_{KR}}\right)}{\sqrt{1+\frac{\mu\alpha_{(5)}\Phi^{2}_{KR}}{Q^{2}_{KR}}
+\frac{\alpha_{(5)}}{M^{5}_{(5)}}
\left(\frac{4\Lambda_{(5)}}{3}-\frac{128\Phi^{3}_{KR}M^{3}_{(5)}}{Q_{KR}}\right)}}
\right.\\ \left.~~~~~~~~~~~~~~~~~~~~~~~~~~~~~~~~~~~~~~~~~~~~~~~~~~~~~~~~~~~~~~~~~~~~~~~~~~~~~~~~~~~~~~~~~~~~~
-\frac{\left(\frac{768\Phi^{\frac{7}{2}}_{KR}\alpha^{2}_{(5)}}{M^{2}_{(5)}Q^{\frac{5}{2}}_{KR}}-\frac{4\mu\alpha^{2}_{(5)}\Phi^{\frac{5}{2}}}{Q^{\frac{7}{2}}_{KR}}
\right)\left[\mu-\frac{192\Phi_{KR}Q_{KR}}{M^{2}_{(5)}}\right]
}{\left[1+\frac{\mu\alpha_{(5)}\Phi^{2}_{KR}}{Q^{2}_{KR}}
+\frac{\alpha_{(5)}}{M^{5}_{(5)}}
\left(\frac{4\Lambda_{(5)}}{3}-\frac{128\Phi^{3}_{KR}M^{3}_{(5)}}{Q_{KR}}\right)\right]^{\frac{3}{2}}}\right\}.\end{array}\ee

\begin{figure}[htb]
{\centerline{\includegraphics[width=12cm, height=8cm] {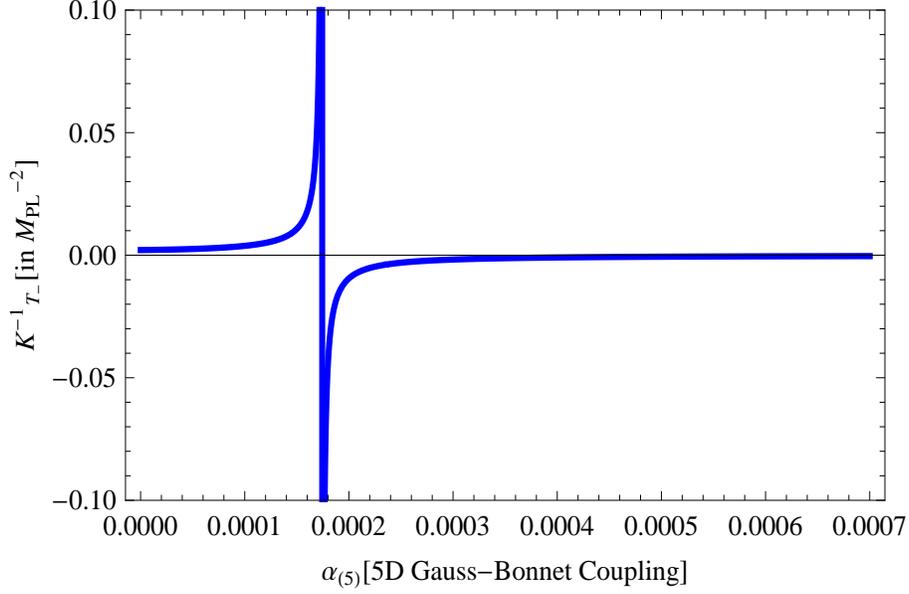}}}
\caption{Variation
 of inverse of isothermal compressibility $K^{-1}_{T_{-}}$ 
with five dimensional Gauss-Bonnet coupling $\alpha_{(5)}$  Kalb Ramond charge $Q_{KR}=10^{-4}$.
 Here we use ADM mass parameter $\mu=0.028$ in the Planckian unit.} \label{fig6}
\end{figure}

\begin{figure}[htb]
{\centerline{\includegraphics[width=12cm, height=8cm] {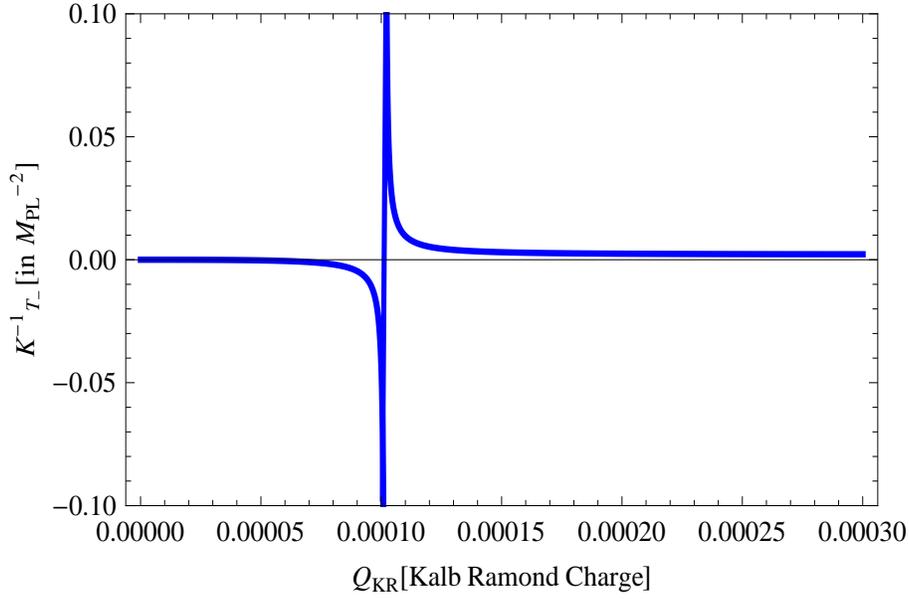}}}
\caption{Variation
 of inverse of isothermal compressibility $K^{-1}_{T_{-}}$ 
with Kalb Ramond charge $Q_{KR}$ five dimensional Gauss-Bonnet coupling $\alpha_{(5)}=0.00018$. 
 Here we use ADM mass parameter $\mu=0.028$ in the Planckian unit.} \label{fig6a}
\end{figure}

In the asymptotic limit $\alpha_{(5)}\rightarrow 0$, the isothermal compressibility simplifies to:
\be\begin{array}{llll}\label{q1}
    \displaystyle \lim_{\alpha_{(5)}\rightarrow 0}K^{-1}_{T_{-}}:=K^{-1}_{T_{\bf GR}}=-Q_{KR}\frac{\left[\frac{32\Phi^{\frac{5}{2}}_{\star}}{Q^{\frac{3}{2}}_{KR}}
-\frac{3\mu M^{2}_{(5)}\Phi^{\frac{3}{2}}_{\star}
}{8Q^{\frac{5}{2}}_{KR}}
-\frac{\Lambda_{(5)}}{6M^{3}_{(5)}\sqrt{Q_{KR}\Phi_{\star}}}\right]}{\left[\frac{3\mu M^{2}_{(5)}\sqrt{\Phi_{\star}}
}{8Q^{\frac{3}{2}}_{KR}}
+\frac{\Lambda_{(5)}\sqrt{Q_{KR}}}{6M^{3}_{(5)}\Phi^{\frac{3}{2}}_{\star}}-\frac{160\Phi^{\frac{3}{2}}_{\star}}{\sqrt{Q_{KR}}}\right]}
   \end{array}\ee

where $\Phi_{\star}=\Phi(r=r_{\star})$. In figure(\ref{fig6}) we have shown the behaviour of the inverse of the isothermal
compressibility with respect to the five dimensional Gauss-Bonnet coupling $\alpha_{(5)}$. The discontinuity appearing in the plot 
has clearly shown the existence of phase transition in our set up. We have also shown the behaviour of the inverse of the isothermal
compressibility with respect to the Kalb Ramond charge for a fixed value of $\alpha_{(5)}=0.00018$ in figure(\ref{fig6a}).
\subsection{\bf Volume expansion coefficient or volume expansivity}
In the context of black hole thermodynamics the volume expansivity or volume expansion coefficient is defined as:
\be\label{c1}
\beta^{-1}_{Q_{KR}}=Q_{KR}
\left(\frac{\partial T_{-}}{\partial Q_{KR}}\right)_{ \Phi_{KR}}.\ee
Now using equation(\ref{iou31}) the volume expansivity for the Kalb Ramond black hole can be computed as:
\be\begin{array}{llll}\label{c21}
    \displaystyle \beta^{-1}_{Q_{KR}}=Q_{KR}\Xi(\alpha_{(5)},Q_{KR},\Phi_{KR},\Lambda_{(5)},\mu)
   \end{array}\ee

\begin{figure}[htb]
{\centerline{\includegraphics[width=12.5cm, height=9.75cm] {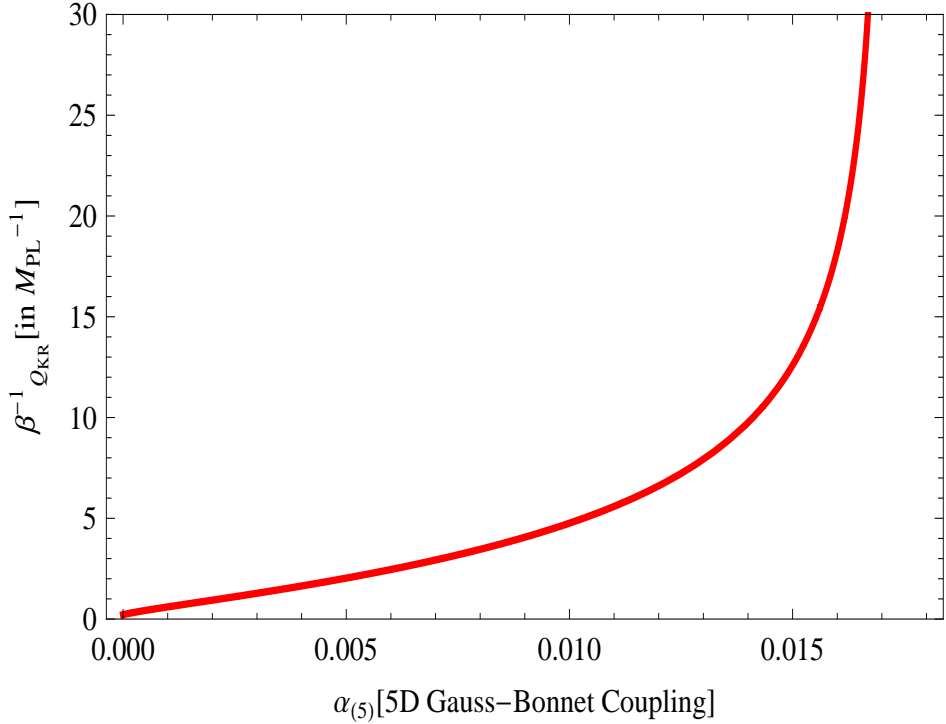}}}
\caption{Variation
 of inverse of volume expansivity $\beta^{-1}_{Q_{KR}}$ 
with five dimensional Gauss-Bonnet coupling $\alpha_{(5)}$ Kalb Ramond charge $Q_{KR}=10^{-4}$.
 Here we use ADM mass parameter $\mu=0.028$ in the Planckian unit.} \label{fig7}
\end{figure}

\begin{figure}[htb]
{\centerline{\includegraphics[width=12.5cm, height=9.75cm] {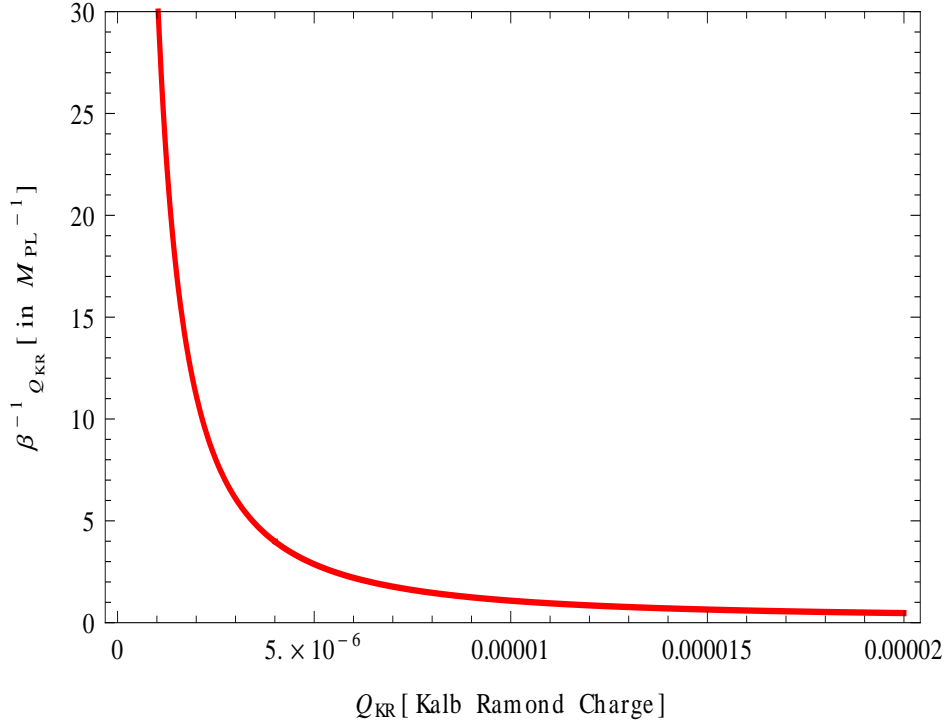}}}
\caption{Variation
 of inverse of volume expansivity $\beta^{-1}_{Q_{KR}}$ 
with Kalb Ramond charge $Q_{KR}$ for five dimensional Gauss-Bonnet coupling $\alpha_{(5)}=0.001$.
 Here we use ADM mass parameter $\mu=0.028$ in the Planckian unit system.} \label{fig7a}
\end{figure}
where the expression for $\Xi(\alpha_{(5)},Q_{KR},\Phi_{KR},\Lambda_{(5)},\mu)$ is mentioned in equation(\ref{b1}).
Specifically in the asymptotic limit $\alpha_{(5)}\rightarrow 0$, the volume expansivity reduces to:
\be\begin{array}{llll}\label{km1}
    \displaystyle \lim_{\alpha_{(5)}\rightarrow 0}\beta^{-1}_{Q_{KR}}=\beta^{-1}_{\bf GR}=\frac{1}{4\pi}
\left[\frac{32\Phi^{\frac{5}{2}}_{\star}}{\sqrt{Q_{KR}}}
-\frac{3\mu M^{2}_{(5)}\Phi^{\frac{3}{2}}_{\star}
}{8Q^{\frac{3}{2}}_{KR}}
-\frac{\Lambda_{(5)}}{6M^{3}_{(5)}}\sqrt{\frac{Q_{KR}}{\Phi_{\star}}}\right].
   \end{array}\ee
In figure(\ref{fig7}) we have shown the behaviour of the inverse of the volume expansivity with respect to the five dimensional Gauss-Bonnet coupling $\alpha_{(5)}$.
It is evident from the plot that, as the numerical value of the five dimensional Gauss-Bonnet coupling
 increases, the corresponding value of the volume expansivity decreases. We have also shown the behaviour of the inverse of the volume expansivity
 with respect to the Kalb Ramond charge for a fixed value of $\alpha_{(5)}=0.001$ in figure(\ref{fig7a}).

\subsection{\bf Specific heat at constant Kalb Ramond electric potential}
In this context the specific heat at constant Kalb Ramond electric potential is defined as:
\be\label{sp1}
C_{\Phi_{KR}}=T_{-}\left(\frac{\partial S_{H}}{\partial T_{-}}\right)_{\Phi_{KR}}=T_{-}\frac{\left(\frac{\partial S_{H}}{\partial r_{H}}\right)_{\Phi_{KR}}}
{\left(\frac{\partial T_{-}}{\partial r_{H}}\right)_{\Phi_{KR}}}
\ee
which plays the analogous role of specific heat at constant pressure ($C_{P}$) in usual equilibrium thermodynamics.
To serve this purpose here we have to express  
equation(\ref{iou3}) and equation(\ref{kl23}) in terms of the Kalb Ramond potential ($\Phi_{KR}$) by eliminating the Kalb Ramond charge ($Q_{KR}$) using the relation 
$\Phi_{KR}=\frac{Q_{KR}}{r^{2}_{H}}$.
Consequently we have
\be\begin{array}{llll}\label{iou36}
    \displaystyle T_{-}=- \frac{M^{2}_{(5)}}{16\pi\alpha_{(5)}}\left\{2r_{H}\sqrt{1+\frac{\mu\alpha_{(5)}}{r^{4}_{H}}+\frac{\alpha_{(5)}}{M^{5}_{(5)}}
\left(\frac{4\Lambda_{(5)}}{3}-\frac{128\Phi^{2}_{KR}M^{3}_{(5)}}{r^{2}_{H}}\right)}
+\frac{\left(\frac{768\Phi^{2}_{KR}\alpha_{(5)}}{M^{2}_{(5)}r_{H}}-\frac{4\mu\alpha_{(5)}}{r^{3}_{H}}\right)}{\sqrt{1+\frac{\mu\alpha_{(5)}}{r^{4}_{H}}+\frac{\alpha_{(5)}}{M^{5}_{(5)}}
\left(\frac{4\Lambda_{(5)}}{3}-\frac{128\Phi^{2}_{KR}M^{3}_{(5)}}{r^{2}_{H}}\right)}}\right\}
   \end{array}\ee
Using equation(\ref{iou36}), the expression for the specific heat turns out to be:
\be\begin{array}{lllll}\label{a1a}
 C^{-}_{\Phi_{KR}}=\frac{\left[\frac{3}{2}\pi^{2} r^{2}_{H}
+3\alpha_{(5)}\pi^2\right]\left\{2r_{H}\sqrt{1+\frac{\mu\alpha_{(5)}}{r^{4}_{H}}+\frac{\alpha_{(5)}}{M^{5}_{(5)}}
\left(\frac{4\Lambda_{(5)}}{3}-\frac{128\Phi^{2}_{KR}M^{3}_{(5)}}{r^{2}_{H}}\right)}
+\frac{\left(\frac{768\Phi^{2}_{KR}\alpha_{(5)}}{M^{2}_{(5)}r_{H}}-\frac{4\mu\alpha_{(5)}}{r^{3}_{H}}\right)}{\sqrt{1+\frac{\mu\alpha_{(5)}}{r^{4}_{H}}+\frac{\alpha_{(5)}}{M^{5}_{(5)}}
\left(\frac{4\Lambda_{(5)}}{3}-\frac{128\Phi^{2}_{KR}M^{3}_{(5)}}{r^{2}_{H}}\right)}}\right\}}
{\left\{2\sqrt{1+\frac{\mu\alpha_{(5)}}{r^{4}_{H}}+\frac{\alpha_{(5)}}{M^{5}_{(5)}}
\left(\frac{4\Lambda_{(5)}}{3}-\frac{128\Phi^{2}_{KR}M^{3}_{(5)}}{r^{2}_{H}}\right)}
-\frac{\left(\frac{512\Phi^{2}_{KR}\alpha_{(5)}}{M^{2}_{(5)}r^{2}_{H}}-\frac{8\mu\alpha_{(5)}}{r^{4}_{H}}\right)}{\sqrt{1+\frac{\mu\alpha_{(5)}}{r^{4}_{H}}+\frac{\alpha_{(5)}}{M^{5}_{(5)}}
\left(\frac{4\Lambda_{(5)}}{3}-\frac{128\Phi^{2}_{KR}M^{3}_{(5)}}{r^{2}_{H}}\right)}}-
\frac{\left(\frac{768\Phi^{2}_{KR}\alpha_{(5)}}{M^{2}_{(5)}r^{2}_{H}}-\frac{4\mu\alpha_{(5)}}{r^{3}_{H}}\right)^{2}}
{\left[1+\frac{\mu\alpha_{(5)}}{r^{4}_{H}}+\frac{\alpha_{(5)}}{M^{5}_{(5)}}
\left(\frac{4\Lambda_{(5)}}{3}-\frac{128\Phi^{2}_{KR}M^{3}_{(5)}}{r^{2}_{H}}\right)\right]^{\frac{3}{2}}}\right\}}
   \end{array}\ee

\begin{figure}[htb]
{\centerline{\includegraphics[width=12cm, height=8cm] {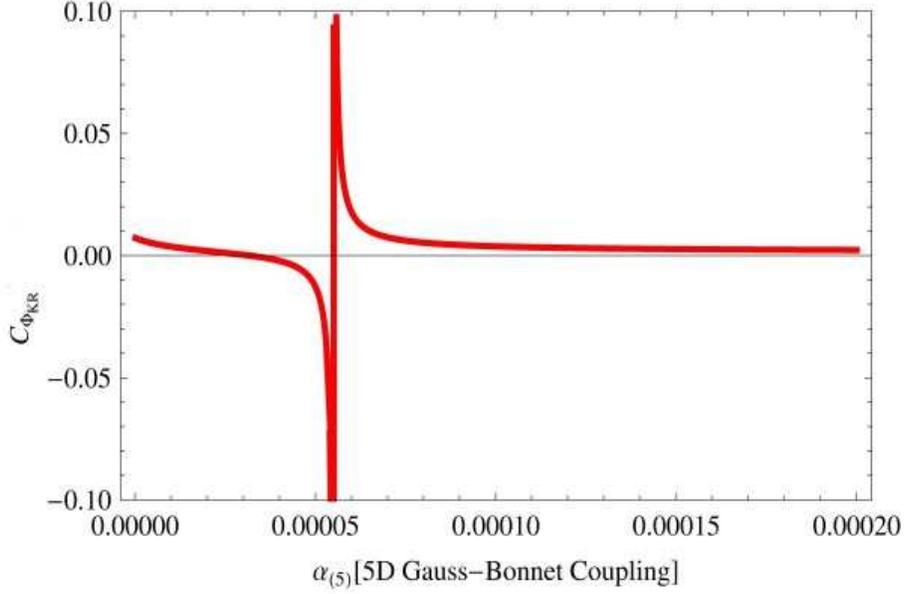}}}
\caption{Variation
 of specific heat $C^{-}_{\Phi_{KR}}(:=C_{\Phi_{KR}})$ 
with five dimensional Gauss-Bonnet coupling $\alpha_{(5)}$  for $\Phi_{KR}=2\times 10^{-3},\Lambda_{(5)}<0$.
 Here we use the ADM mass parameter $\mu=0.028$ in the Planckian unit.} \label{fig8}
\end{figure}

\begin{figure}[htb]
{\centerline{\includegraphics[width=12cm, height=8cm] {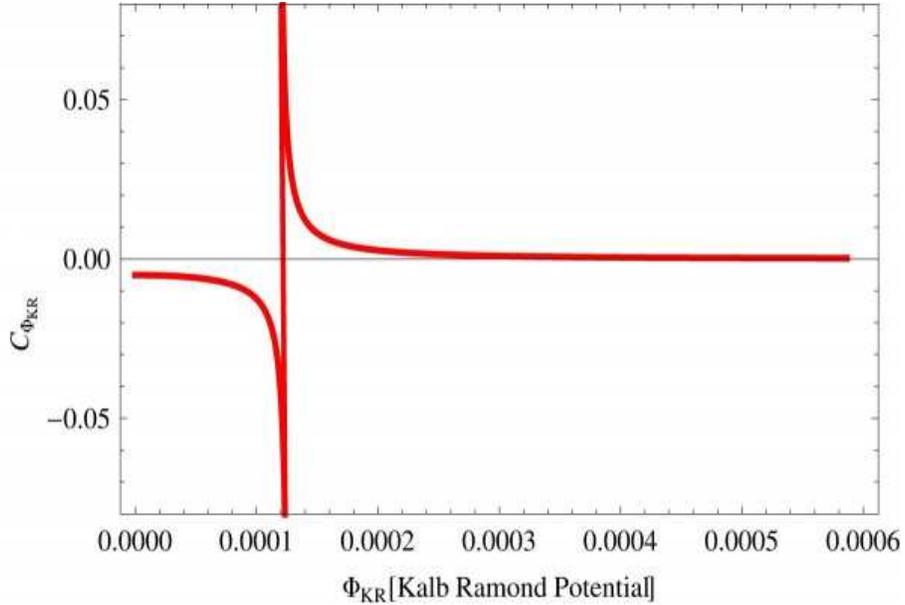}}}
\caption{Variation
 of specific heat $C^{-}_{\Phi_{KR}}(:=C_{\Phi_{KR}})$ 
with Kalb Ramond potential $\Phi_{KR}$ for five dimensional Gauss-Bonnet coupling $\alpha_{(5)}=\alpha^{Crit}_{(5)}$.
 Here we use the ADM mass parameter $\mu=0.028$ in the Planckian unit.} \label{fig8a}
\end{figure}

In the asymptotic limit $\alpha_{(5)}\rightarrow 0$, equation(\ref{a1a}) reduces to the following expression:
\be\begin{array}{llll}\label{cv1}
    \displaystyle \lim_{\alpha_{(5)}\rightarrow 0}C^{-}_{\Phi_{\star}}=C^{\bf GR}_{\Phi_{\star}}=\frac{\left[\frac{3}{2}\pi^{2} r^{2}_{\star}
+3\alpha_{(5)}\pi^2\right]
\left[\frac{\mu M^{2}_{(5)}}{4r^{3}_{\star}}
-\frac{\Lambda_{(5)}r_{\star}}{3M^{3}_{(5)}}-\frac{64\Phi^{2}_{\star}}{r_{\star}}\right]}{\left[\frac{64\Phi^{2}_{\star}}{r^{2}_{\star}}-\frac{3\mu M^{2}_{(5)}}{4r^{4}_{\star}}
-\frac{\Lambda_{(5)}}{3M^{3}_{(5)}}\right]}.
   \end{array}\ee

In $\Phi_{KR}\rightarrow 0$ limit the expression for the specific heat reduces to equation(\ref{g1}).
In figure(\ref{fig8}) we have explicitly shown the behaviour of specific heat at constant Kalb Ramond potential $C^{-}_{\Phi_{KR}}$
with respect to the five dimensional Gauss-Bonnet coupling $\alpha_{(5)}$.
 The discontinuity appearing in this plot directly confirms the appearance of phase transition in the present context.
Additionally we have also shown the behaviour of the inverse of the specific heat $C^{-}_{\Phi_{KR}}$
 with respect to the Kalb Ramond charge for a fixed value of $\alpha_{(5)}=\alpha^{Crit}_{(5)}$ in figure(\ref{fig8a}).

\subsection{\bf Legendre transformation and free energy}

In presence of Gauss-Bonnet coupling ($\alpha_{(5)}$) for Kalb Ramond black hole, the ``Gibbs free energy'' is defined as`:
\be\label{dq1}
G_{KR}=M_{KR}-T_{-}S_{H}-\Phi_{KR}Q_{KR}=G_{KR}(T_{-},\Phi_{KR})\ee
where the black hole mass ($M_{KR}$) plays the analogous role of internal energy ($U$) of a thermodynamical system.
Taking infinitesimal reversible change in the ``Gibbs free energy'' and using the {\it first} and {\it second} law of black hole thermodynamics, we get:
\be\begin{array}{llll}\label{zx1}
    \displaystyle dG_{KR}=\left(\frac{\partial G_{KR}}{\partial T_{-}}\right)_{\Phi_{KR}}dT_{-}
+\left(\frac{\partial G_{KR}}{\partial \Phi_{KR}}\right)_{T_{-}}d\Phi_{KR}=-S_{H}dT_{-}-Q_{KR}d\Phi_{KR}
   \end{array}\ee
This implies 
\be\begin{array}{llll}\label{cz1}
    \displaystyle S_{H}=-\left(\frac{\partial G_{KR}}{\partial T_{-}}\right)_{\Phi_{KR}}, \\
\displaystyle Q_{KR}=-\left(\frac{\partial G_{KR}}{\partial \Phi_{KR}}\right)_{T_{-}}.
   \end{array}\ee
Now using the Legendre transformation on ``Gibbs free energy'' it is possible to construct the other free energies like ``Enthalpy''($H_{KR}$),
 ``Helmholtz free energy''($F_{KR}$)
as well as ``black hole mass''($M_{KR}$). Let us start with $G_{KR}\rightarrow H_{KR}$ transformation which gives
\be\begin{array}{llll}\label{n1}
    \displaystyle H_{KR}=G_{KR}-\left(\frac{\partial G_{KR}}{\partial T_{-}}\right)_{\Phi_{KR}}T_{-}=M_{KR}-\Phi_{KR}Q_{KR}=H_{KR}(S_{H},\Phi_{KR})
   \end{array}\ee
Taking infinitesimal reversible change in the ``Enthalpy'' and using the {\it first} and {\it second} law of black hole thermodynamics we get:
\be\begin{array}{llll}\label{zz1}
    \displaystyle dH_{KR}=\left(\frac{\partial H_{KR}}{\partial S_{H}}\right)_{\Phi_{KR}}dS_{H}
+\left(\frac{\partial H_{KR}}{\partial \Phi_{KR}}\right)_{S_{H}}d\Phi_{KR}=T_{H}dS_{H}-Q_{KR}d\Phi_{KR}
   \end{array}\ee
This implies 
\be\begin{array}{llll}\label{cz1a}
    \displaystyle T_{H}=\left(\frac{\partial H_{KR}}{\partial S_{H}}\right)_{\Phi_{KR}}, \\
\displaystyle Q_{KR}=-\left(\frac{\partial H_{KR}}{\partial \Phi_{KR}}\right)_{T_{-}}.
   \end{array}\ee
Next we consider $G_{KR}\rightarrow F_{KR}$ transformation which gives
\be\begin{array}{llll}\label{n1aa}
    \displaystyle F_{KR}=G_{KR}-\left(\frac{\partial G_{KR}}{\partial \Phi_{KR}}\right)_{T_{-}}\Phi_{KR}=M_{KR}-T_{-}S_{H}=F_{KR}(T_{-},Q_{KR})
   \end{array}\ee
Taking infinitesimal reversible change in the ``Enthalpy'' and using the {\it first} and {\it second} law of black hole thermodynamics we get:
\be\begin{array}{llll}\label{zz1a}
    \displaystyle dF_{KR}=\left(\frac{\partial F_{KR}}{\partial T_{-}}\right)_{Q_{KR}}dT_{-}
+\left(\frac{\partial F_{KR}}{\partial Q_{KR}}\right)_{T_{-}}dQ_{KR}=\Phi_{KR}dQ_{KR}-S_{H}dT_{-}
   \end{array}\ee
This implies 
\be\begin{array}{llll}\label{cz1aa}
    \displaystyle S_{H}=-\left(\frac{\partial F_{KR}}{\partial T_{-}}\right)_{Q_{KR}}, \\
\displaystyle \Phi_{KR}=\left(\frac{\partial F_{KR}}{\partial Q_{KR}}\right)_{T_{-}}.
   \end{array}\ee
At last considering $G_{KR}\rightarrow M_{KR}$ transformation the Kalb Ramond black hole ADM mass can be written as:
\be\begin{array}{llll}\label{n1a3}
    \displaystyle M_{KR}=G_{KR}-\left(\frac{\partial G_{KR}}{\partial \Phi_{KR}}\right)_{T_{-}}\Phi_{KR}
-\left(\frac{\partial G_{KR}}{\partial T_{-}}\right)_{\Phi_{KR}}T_{-}=M_{KR}-T_{-}S_{H}=M_{KR}(S_{H},Q_{KR})
   \end{array}\ee
Taking infinitesimal reversible change in the ``ADM mass'' of black hole we get:
\be\begin{array}{llll}\label{zz16}
    \displaystyle dM_{KR}=\left(\frac{\partial M_{KR}}{\partial S_{H}}\right)_{Q_{KR}}dS_{H}
+\left(\frac{\partial M_{KR}}{\partial Q_{KR}}\right)_{S_{H}}dQ_{KR}=\Phi_{KR}dQ_{KR}+T_{-}dS_{H}
   \end{array}\ee
which is the well known {\it first law of black hole thermodynamics}. This implies 
\be\begin{array}{llll}\label{cz1a6}
    \displaystyle T_{-}=\left(\frac{\partial M_{KR}}{\partial S_{H}}\right)_{Q_{KR}}, \\
\displaystyle \Phi_{KR}=\left(\frac{\partial M_{KR}}{\partial Q_{KR}}\right)_{S_{H}}.
   \end{array}\ee
\section{\bf Confirmatory test of second order phase transition via Ehrenfest's Theorem}
In the present section we focus on the second order phase transition phenomena in Kalb Ramond AdS blackholes.
For that purpose here we use all black hole thermodynamic parameters discussed in the earlier sections.

\subsection{\bf Thermodynamic Maxwell's Relations}
To establish the thermodynamic Maxwell's relations here we remember that all the Kalb Ramond black hole thermodynamic potentials
$G_{KR}$, $H_{KR}$, $F_{KR}$ and $M_{KR}$ are thermodynamic state functions and hence any infinitesimal reversible change in the thermodynamic
potentials are exact differentials. Consequently we have:

\be\begin{array}{llll}\label{cz1x}
    \displaystyle \left(\frac{\partial^{2} G_{KR}}{\partial 
\Phi_{KR}\partial T_{-}}\right)=\left(\frac{\partial^{2} G_{KR}}{\partial T_{-}\partial \Phi_{KR}}\right)
\Rightarrow -\left(\frac{\partial S_{H}}{\partial 
\Phi_{KR}}\right)_{T_{-}}=-\left(\frac{\partial Q_{KR}}{\partial T_{-}}\right)_{\Phi_{KR}}
   \end{array}\ee

\be\begin{array}{llll}\label{cz1x1}
    \displaystyle \left(\frac{\partial^{2} H_{KR}}{\partial 
\Phi_{KR}\partial S_{H}}\right)=\left(\frac{\partial^{2} H_{KR}}{\partial S_{H}\partial \Phi_{KR}}\right)
\Rightarrow -\left(\frac{\partial T_{-}}{\partial 
\Phi_{KR}}\right)_{S_{H}}=\left(\frac{\partial Q_{KR}}{\partial S_{H}}\right)_{\Phi_{KR}}
   \end{array}\ee

\be\begin{array}{llll}\label{cz1x11}
    \displaystyle \left(\frac{\partial^{2} F_{KR}}{\partial 
Q_{KR}\partial T_{-}}\right)=\left(\frac{\partial^{2} F_{KR}}{\partial T_{-}\partial Q_{KR}}\right)
\Rightarrow -\left(\frac{\partial S_{H}}{\partial 
Q_{KR}}\right)_{T_{-}}=-\left(\frac{\partial \Phi_{KR}}{\partial T_{-}}\right)_{Q_{KR}}
   \end{array}\ee

\be\begin{array}{llll}\label{cz1x111}
    \displaystyle \left(\frac{\partial^{2} M_{KR}}{\partial 
Q_{KR}\partial S_{H}}\right)=\left(\frac{\partial^{2} M_{KR}}{\partial S_{H}\partial Q_{KR}}\right)
\Rightarrow -\left(\frac{\partial T_{-}}{\partial 
Q_{KR}}\right)_{S_{H}}=-\left(\frac{\partial \Phi_{KR}}{\partial S_{H}}\right)_{Q_{KR}}.
   \end{array}\ee
Here equation(\ref{cz1x}-\ref{cz1x111}) are known as the thermodynamic Maxwell's relations which are universally satisfied by 
the charged Kalb Ramond black hole as introduced in this article. 


\subsection{\bf Verification of Ehrenfest's Theorem and Prigogine-Defay ratio}

From the basic understanding of statistical mechanics it is a well established fact that 
discontinuity in the heat capacity does not always imply a second order phase transition, moreover it
implies a continuous higher order phase transition in general. In this context the master equations namely the Ehrenfest’s equations play a crucial
role in order to determine the behaviour of the higher order phase transitions for various conventional
thermodynamical systems. Additionally such technique can be very easily applied to the 
various thermodynamical systems from which the nature of the corresponding phase transition can also be
determined. On the contrary, if a phase transition is not at all a second
order type then usually {\it Prigogine-Defay (PD)
ratio} \cite{diba3,thm,diba4,thm1,jackle,modak,modak2,modak3} is used to measure the degree of its deviation from the second order phase transition.
 Following synonymous approach in the context of charged Kalb Ramond black hole, we classify the phase
transition phenomena in the context of equilibrium black holes thermodynamics and examine the applicability of 
Ehrenfest's tool for
charged Kalb Ramond black hole.

In conventional equilibrium thermodynamics, the first and the second Ehrenfest’s equations can be written as:
\be\begin{array}{llll}\label{ern1}
    \displaystyle \left(\frac{\partial P}{\partial T}\right)_{S}=\frac{1}{VT}\frac{\left(C_{P_{2}}-C_{P_{1}}\right)}{\left(\beta_{V_{2}}-\beta_{V_{1}}\right)}
=\frac{\Delta C_{P}}{VT\Delta\beta_{V}},\end{array}\ee
\be\begin{array}{llll}\label{ern2}
\displaystyle \left(\frac{\partial P}{\partial T}\right)_{V}=\frac{\left(\beta_{V_{2}}-\beta_{V_{1}}\right)}{\left(K_{T_{2}}-K_{T_{1}}\right)}
=\frac{\Delta\beta_{V}}{\Delta K_{T}}.
   \end{array}\ee
In the context of charged Kalb Ramond black hole thermodynamics equation(\ref{ern1})
and equation(\ref{ern2}) rewritten as:
\be\begin{array}{llll}\label{ern1a}
    \displaystyle -\left(\frac{\partial \Phi_{KR}}{\partial T_{-}}\right)_{S_{H}}=\frac{1}{Q_{KR}T_{-}}
\frac{\left(C_{\Phi_{KR2}}-C_{\Phi_{KR1}}\right)}{\left(\beta_{Q_{KR2}}-\beta_{Q_{KR1}}\right)}
=\frac{\Delta C_{\Phi_{KR}}}{Q_{KR}T_{-}\Delta\beta_{Q_{KR}}},\end{array}\ee
\be\begin{array}{llll}\label{ern2b}
\displaystyle -\left(\frac{\partial \Phi_{KR}}{\partial T_{-}}\right)_{Q_{KR}}
=\frac{\left(\beta_{Q_{KR2}}-\beta_{Q_{KR1}}\right)}{\left(K_{T_{-2}}-K_{T_{-1}}\right)}
=\frac{\Delta\beta_{Q_{KR}}}{\Delta K_{T_{-}}}.
   \end{array}\ee
Throughout the analysis the suffices ``1'' and ``2'' represents two distinct phases of the black hole thermodynamical system. 
Moreover each of the symbols and their expressions are computed in the earlier sections.
To examine the applicability of equation(\ref{ern1})
and equation(\ref{ern2}) here we start with the definition of volume expansivity mentioned in equation(\ref{c1}).
This gives:
\be\begin{array}{llll}\label{bnq1}
    \displaystyle Q_{KR}\beta_{Q_{KR}}=\left(\frac{\partial Q_{KR}}{\partial T_{-}}\right)_{\Phi_{KR}}=\frac{C_{\Phi_{KR}}}{T_{-}}
\left(\frac{\partial Q_{KR}}{\partial T_{-}}\right)_{\Phi_{KR}}.
   \end{array}\ee
After establishing the existence of the two Ehrenfest’s equations for known volume expansivity and isothermal compressibility here our prime
objective is to determine the order of the phase transition. For this, we shall analytically check the
applicability of the two Ehrenfest's equations at the points of discontinuity $S_{iH}\forall i$. 
In the context of equilibrium statistical mechanics such points are identified to be the ``critical points'' or ``transition points''.

   Now taking infinitesimal reversible change in both sides of equation(\ref{bnq1}) we get:
\be\begin{array}{llll}\label{h1}
    \displaystyle \frac{\Delta C_{\Phi_{KR}}}{T_{i-}Q_{iKR}\Delta\beta_{Q_{iKR}}}=\left[\left(\frac{\partial T_{-}}{\partial Q_{KR}}\right)_{\Phi_{KR}}\right]_{S_{H}=S_{iH}}
\\ \displaystyle~~~~~~~~~~~~~~~~~~~~~~~=-\left[\left(\frac{\partial S_{H}}{\partial T_{-}}\right)_{Q_{KR}}
\left(\frac{\partial T_{-}}{\partial Q_{KR}}\right)_{\Phi_{KR}}\right]_{S_{H}=S_{iH}}
\\  \displaystyle~~~~~~~~~~~~~~~~~~~~~~~=-\left[\frac{\Xi(\alpha_{(5)},Q_{KR},\Phi_{KR},\Lambda_{(5)},\mu)}
{\Upsilon(\alpha_{(5)},Q_{KR},\Phi_{KR},\Lambda_{(5)},\mu)}\sqrt{\frac{Q_{KR}}{\Phi_{KR}}}\right]_{S_{H}=S_{iH}} 
\end{array}\ee
where we use the well known thermodynamical identity:
\be\begin{array}{llll}\label{ver1}
    \displaystyle \left(\frac{\partial S_{H}}{\partial T_{-}}\right)_{Q_{KR}}
\left(\frac{\partial T_{-}}{\partial Q_{KR}}\right)_{\Phi_{KR}}\left(\frac{\partial Q_{KR}}{\partial T_{-}}\right)_{\Phi_{KR}}=-1.
   \end{array}\ee
Here we define a new function:
\be\begin{array}{llll}\label{ver2}
   \tiny \Upsilon(\alpha_{(5)},Q_{KR},\Phi_{KR},\Lambda_{(5)},\mu):=\left\{2\sqrt{1+\frac{\mu\alpha_{(5)}}{r^{4}_{H}}+\frac{\alpha_{(5)}}{M^{5}_{(5)}}
\left(\frac{4\Lambda_{(5)}}{3}-\frac{128Q^{2}_{KR}M^{3}_{(5)}}{r^{6}_{H}}\right)}
-\frac{\left(\frac{3072Q^{2}_{KR}\alpha_{(5)}}{M^{2}_{(5)}r^{6}_{H}}-\frac{8\mu\alpha_{(5)}}{r^{4}_{H}}\right)}
{\sqrt{1+\frac{\mu\alpha_{(5)}}{r^{4}_{H}}+\frac{\alpha_{(5)}}{M^{5}_{(5)}}
\left(\frac{4\Lambda_{(5)}}{3}-\frac{128Q^{2}_{KR}M^{3}_{(5)}}{r^{6}_{H}}\right)}}\right.\\ \left.~~~~~~~~~~~~~~~~~~~~
~~~~~~~~~~~~~~~~~~~~~~~~~~~~~~~~~~~~~~~~~~~~~~~~~~~~~~~~~~~~~~~~~~~~~-
\frac{\left(\frac{768Q^{2}_{KR}\alpha_{(5)}}{M^{2}_{(5)}r^{5}_{H}}-\frac{4\mu\alpha_{(5)}}{r^{3}_{H}}\right)^{2}
}
{2r^{2}_{H}\left[1+\frac{\mu\alpha_{(5)}}{r^{4}_{H}}+\frac{\alpha_{(5)}}{M^{5}_{(5)}}
\left(\frac{4\Lambda_{(5)}}{3}-\frac{128Q^{2}_{KR}M^{3}_{(5)}}{r^{6}_{H}}\right)\right]^{\frac{3}{2}}}\right\}.
   \end{array}\ee

On the other hand 
\be\begin{array}{lll}\label{lz1}
    \displaystyle \left[\left(\frac{\partial \Phi_{KR}}{\partial T_{-}}\right)_{S_{H}}\right]_{S_{H}=S_{iH}}=
\left[\frac{\Xi(\alpha_{(5)},Q_{KR},\Phi_{KR},\Lambda_{(5)},\mu)}
{\Upsilon(\alpha_{(5)},Q_{KR},\Phi_{KR},\Lambda_{(5)},\mu)}\sqrt{\frac{Q_{KR}}{\Phi_{KR}}}\right]_{S_{H}=S_{iH}}. 
   \end{array}\ee
This implies
\be\begin{array}{llll}\label{we1}
    \displaystyle \frac{\Delta C_{\Phi_{iKR}}}{T_{i-}Q_{iKR}\Delta\beta_{Q_{iKR}}}=-\left[\left(\frac{\partial \Phi_{KR}}{\partial T_{-}}\right)_{S_{H}}\right]_{S_{H}=S_{iH}}
   \end{array}\ee
i.e. the first Ehrenfest's theorem is established for charged Kalb Ramond black hole.

In order to evaluate the left hand side of the second Ehrenfest's equation, we use the thermodynamical expression of temperature as:
\be\label{dep1}
T_{-}=T_{-}(S_{H},\Phi_{KR})\ee
which essentially gives
\be\begin{array}{lllll}\label{u1}
    \displaystyle \left(\frac{\partial T_{-}}{\partial \Phi_{KR}}\right)_{Q_{KR}}=
\left(\frac{\partial T_{-}}{\partial S_{H}}\right)_{\Phi_{KR}}\left(\frac{\partial S_{H}}{\partial\Phi_{KR} }\right)_{Q_{KR}}+\left(\frac{\partial T_{-}}{\partial\Phi_{KR}}\right)_{S_{H}}.
   \end{array}\ee
Now from the definition of the critical point it is obvious that 
\be\begin{array}{llll}\label{jsd1}
    \displaystyle \left[\left(\frac{\partial T_{-}}{\partial S_{H}}\right)_{\Phi_{KR}}\right]_{S_{H}=S_{iH}}=0.
\end{array}\ee
This implies 
\be\begin{array}{llll}\label{afg1}
    \displaystyle -\left[\left(\frac{\partial \Phi_{KR}}{\partial T_{-}}\right)_{Q_{KR}}\right]_{S_{H}=S_{iH}}=
-\left[\left(\frac{\partial \Phi_{KR}}{\partial T_{-}}\right)_{S_{H}}\right]_{S_{H}=S_{iH}}=\frac{\Delta C_{\Phi_{iKR}}}{T_{i-}Q_{iKR}\Delta\beta_{Q_{iKR}}}.
   \end{array}\ee
Using the definition of isothermal compressibility from equation(\ref{com1}) we get:
\be\begin{array}{llll}\label{qso1a}
    \displaystyle Q_{KR}K_{i-}=\left(\frac{\partial Q_{KR}}{\partial \Phi_{KR}}\right)_{T_{-}}\\
 \displaystyle~~~~~~~~~~~~=\left(\frac{\partial T_{-}}{\partial \Phi_{KR}}\right)_{Q_{KR}}\left(\frac{\partial Q_{KR}}{\partial T_{-}}\right)_{\Phi_{KR}}
\\  \displaystyle~~~~~~~~~~~~=-Q_{KR}\beta_{Q_{iKR}}\left(\frac{\partial T_{-}}{\partial \Phi_{KR}}\right)_{Q_{KR}}
   \end{array}\ee
where we use the thermodynamic identity:
\be\begin{array}{llll}\label{veri1}
    \displaystyle \left(\frac{\partial Q_{KR}}{\partial \Phi_{KR}}\right)_{T_{-}}
\left(\frac{\partial \Phi_{KR}}{\partial T_{-}}\right)_{Q_{KR}}\left(\frac{\partial T_{-}}{\partial Q_{KR} }\right)_{\Phi_{KR}}=-1.
   \end{array}\ee
Taking infinitesimal reversible change in both the sides of equation(\ref{qso1a}) we get:
\be\begin{array}{llll}\label{veri1}
    \displaystyle \frac{\Delta\beta_{Q_{iKR}}}{\Delta K_{T_{i-}}}=-\left[\left(\frac{\partial \Phi_{KR}}{\partial T_{-}}\right)_{Q_{KR}}\right]_{S_{H}=S_{iH}}
=-\Sigma(\alpha_{(5)},Q_{iKR},\Phi_{iKR},\Lambda_{(5)},\mu)
   \end{array}\ee
which essentially establishes the applicability and existence of second Ehrenfest's equation in our framework. 
Now using equation(\ref{we1}) and equation(\ref{veri1}) the Prigogine-Defay (PD) ratio may be obtained as,
\be\begin{array}{llll}\label{anaw1}
    \displaystyle \Pi_{PD}=\frac{\Delta C_{\Phi_{iKR}}\Delta K_{T_{i-}}}{T_{i-}Q_{iKR}\left(\Delta\beta_{Q_{iKR}}\right)^{2}}
=\left\{-\left[\left(\frac{\partial T_{-}}{\partial \Phi_{KR}}\right)_{S_{H}}\right]_{S_{H}=S_{iH}}\right\}\times
\left\{-\left[\left(\frac{\partial \Phi_{KR}}{\partial T_{-}}\right)_{S_{H}}\right]_{S_{H}=S_{iH}}\right\}=1
   \end{array}\ee
which confirms the existance of second order phase transition in presence of charged Kalb Ramond black hole. 
\section{\bf Critical exponents and behaviour of scaling laws in presence of 5-dimensional Gauss-Bonnet Coupling}

In the earlier sections we have elaborately discussed the detailed features of black hole thermodynamics
and second order phase transition phenomena obtained from charged Kalb Ramond antisymmetric tensor field. In this section our prime focus on
to study the associated critical behaviour of Kalb Ramond black hole by the analysis
with the critical or transition points.
In this context it is very crucial to understand
the behaviour of the divergence appearing in the heat capacity and the singular behaviour of several thermodynamic functions near
the critical point on which the assumptions of equilibrium statistical mechanics are valid.
 For this purpose, here we introduce a set of critical exponents $(\alpha, \beta, \gamma, \delta, \varphi,\psi, \nu, \eta)$
which play a prime character in the context of phase transition and critical phenomena. These critical exponents are
associated with the discontinuities of various kinds of thermodynamical variables. They are to a large
degree universal, depending only on a few fundamental parameters like the dimensionality of
the physical system, symmetry of the order parameter, the range of the interaction, the spin dimension etc. 
Universality is a prediction of the renormalization group theory of phase transitions \cite{golden,ma,luther,nien,beker,calla}, which states that the thermodynamic properties of a system near a phase
 transition are insensitive to the underlying microscopic properties of the system.
These properties of critical exponents are supported by experimental data. The experimental results can be theoretically
 achieved in {\it mean field theory} for higher-dimensional systems ($D\geq 4$). The theoretical treatment
 of lower-dimensional systems ($D=1$ or $D=2$) is more difficult and requires the proper analysis via renormalization group.
In this section without going through the detail of renormalization group flow we estimate the critical exponents for the
various thermodynamic quantities (order parameters) and examine the scaling laws obtained from the antisymmetric Kalb Ramond fields.
 Additionally it is important to mention here that through out the analysis of critical exponents we assume that divergences on the 
correlation length obeys the power law behaviour. But for there are some statistical systems in literature where the power law behaviour 
does not holds good. For example Ising model in $D=2$ shows logarithmic divergence. However, these systems are limiting cases and
 an exception to the rule. Real phase transitions always exhibit power-law behaviour.

\subsection{\bf Critical exponent $\alpha$ }

Let us first start with the analysis of the critical coefficient ``$\alpha$'' associated with the divergence of specific heat $C_{Q_{KR}}$
near the critical point. To serve this purpose the horizon calculated from Kalb Ramond antisymmetric tensor field can be expanded near the critical point 
as:
\be\begin{array}{llll}\label{cr1}
    \displaystyle r_{H}=r_{i}\left(1+\Delta_{C}\right)
   \end{array}\ee
where the subscript ``i'' signifies the number of divergences (critical points) obtained from $C_{Q_{KR}}\rightarrow \infty$. Here $\Delta_{C}<<1$ always.
Let us say the number of positive distinct roots obtained from  $C_{Q_{KR}}\rightarrow \infty$ are ``j'' with the restriction ``$j<i$''. Consequently 
equation(\ref{cr1}) can be written as:
\be\begin{array}{lll}\label{cr2}
    \displaystyle r_{H}=r_{j}\left(1+\Delta_{c}\right).
   \end{array}\ee
Now using the horizon expansion near the critical point stated in equation(\ref{cr2})
temperature for the Kalb Rammond black hole can be expressed as:
\be\begin{array}{lll}\label{cr3}
    \displaystyle T_{-}\left(r_{H}\right)=T_{-}\left(r_{j}\right)\left(1+\epsilon_{C}\right)
   \end{array}\ee
where $\epsilon_{C}<<1$. For fixed Kalb Ramond charge $Q_{KR}$ temperature evaluated at the horizon ($r_{H}$)
can be expanded in a Taylor series around the sufficiently small neighborhood of $r_{j}$ (+ve distinct root)
as:
\be\begin{array}{lll}\label{cr4}
    \displaystyle T_{-}\left(r_{H}\right)=\sum^{\infty}_{n=0}\frac{1}{n!}
\left[\left(\frac{\partial^{n}T_{-}}{\partial r^{n}_{H}}\right)_{Q_{KR}=Q_{C},~r_{H}=r_{j}}\right]\left(r_{H}-r_{j}\right)^{n}.
   \end{array}\ee
Since $C_{Q_{KR}}\rightarrow \infty$  at $r_{H}=r_{j}$ then we have the following constraint:
\be\begin{array}{lll}\label{cr5}
 \displaystyle\left(\frac{\partial T_{-}}{\partial r_{H}}\right)_{Q_{KR}=Q_{C},~r_{H}=r_{j}}=0.
\end{array}\ee
Taking into account this crucial constraint for Kalb Ramond antisymmetric tensor field we get:
\be\begin{array}{lll}\label{cr6}
 \displaystyle |\epsilon_{C}|^{\frac{1}{m}}
=\sqrt[m]{\sum^{\infty}_{n=2}\frac{r^{n}_{j}}{n!T_{-}\left(r_{j}\right)}
\left[\left(\frac{\partial^{n}T_{-}}{\partial r^{n}_{H}}\right)_{Q_{KR}=Q_{C},~r_{H}=r_{j}}\right]\Delta^{n}_{C}}
\end{array}\ee
where $m=2,3,.....\infty$ physically signifies the order of truncation of the the above mentioned Taylor series. 
After some simple algebra from equation(\ref{cr6})
we get:
\be\begin{array}{lll}\label{cr7}
 \displaystyle |T_{-}-T_{j-}|^{\frac{1}{m}}
=\sqrt[m]{\sum^{\infty}_{n=2}\frac{r^{n}_{j}}{n!}
\left[\left(\frac{\partial^{n}T_{-}}{\partial r^{n}_{H}}\right)_{Q_{KR}=Q_{C},~r_{H}=r_{j}}\right]\Delta^{n}_{C}}
\end{array}\ee
where we use the shorthand symbol $T_{-}=T_{-}(r_{H})$ and $T_{j-}=T_{j-}(r_{H})$. Now restricting our analysis 
in the regime of equilibrium statistical mechanics we truncate the Taylor series mentioned in equation(\ref{cr7}) at $m=n=2$.

Let us introduce a function defined as:
\be\begin{array}{llll}\label{cr9a}
    \displaystyle {\bf X}_{2j}=\frac{r^{2}_{j}}{2T_{-}}\left[\left(\frac{\partial^{2}T_{-}}{\partial r^{2}_{H}}\right)_{Q_{KR}=Q_{C},~r_{H}=r_{j}}\right].
   \end{array}\ee
This implies 
\be\begin{array}{lll}\label{cr9}
 \displaystyle \Delta_{C}
=\frac{|\epsilon_{C}|^{\frac{1}{2}}}{\sqrt{{\bf X}_{2j}}}.
\end{array}\ee
Now making use of equation(\ref{cr2}) and equation(\ref{cr9}) in equation(\ref{a1}) we get:
\be\begin{array}{lllll}\label{cr10}
    \displaystyle C^{-}_{Q_{KR}}=\left[\frac{D_{j}}{\sqrt{\epsilon_{C}}}\right]_{r_{H}=r_{k}}
   \end{array}\ee
where at $r_{H}=r_{k}$, $T_{-}(r_{H})>T_{-}(r_{k})$ which implies $\epsilon_{C}>0$. Here always $k<j$ constraint is satisfied.
Additionally $D_{k}$ is the Gauss-Bonnet coupling ($\alpha_{(5)}$) dependent overall constant factor for antisymmetric Kalb Ramond tensor fields
evaluated at the critical point $r_{H}=r_{k}$. It is important to mention here that through out the analysis we are restricting our calculation 
on the linear $\Delta_{C}$ because $\Delta_{C}<<1$. On the other hand for any point close to $r_{H}=r_{l}$ (with $k<l<j$) we have $T_{-}(r_{H}) < T_{-}(r_{l})$
implying that $\epsilon_{C}<0$. Consequently we have:
\be\begin{array}{lllll}\label{cr11}
    \displaystyle C^{-}_{Q_{KR}}=\left[\frac{D_{j}}{\sqrt{-\epsilon_{C}}}\right]_{r_{H}=r_{l}}.
   \end{array}\ee
Combining equation(\ref{cr10}) and equation(\ref{cr11}) the singular behaviour of the specific heat at constant charge near the critical point 
turns out to be:
\be\begin{array}{lllll}\label{cr12}
    \displaystyle C^{-}_{Q_{KR}}=\left[\frac{D_{j}}{\sqrt{|\epsilon_{C}|}}\right]_{r_{H}=r_{j}}\\
\displaystyle~~~~~~~~=\frac{D_{j}T^{\frac{1}{2}}_{j-}}{|T_{-}-T_{j-}|^{\frac{1}{2}}}.
   \end{array}\ee
Comparing equation(\ref{cr12}) with the original power law divergence in the context of statistical mechanics 
\be\begin{array}{lllll}\label{cr13}
    \displaystyle C_{Q}\propto|T-T_{j}|^{-\alpha}
   \end{array}\ee
the critical exponent associated with the singularity in the specific heat at constant charge turns out to be $\alpha=\frac{1}{2}$.

\subsection{\bf Critical exponent $\beta$}
Further we want to determine the critical exponent $\beta$ which is associated to the electric potential ($\Phi$) 
for a fixed value of charge as,
\be\begin{array}{llll}\label{cr14}
    \displaystyle \Phi\left(r_{H}\right)-\Phi\left(r_{j}\right)\propto |T-T_{j}|^{\beta}
   \end{array}\ee
appearing in the context of equilibrium statistical mechanics. To serve this purpose
 we Taylor expand the Kalb Ramond electric potential close to the critical point $r_{H}=r_{j}$ which yields,
\be\begin{array}{llll}\label{cr15}
    \displaystyle \Phi_{KR}\left(r_{H}\right)-\Phi_{KR}\left(r_{j}\right)=\sum^{\infty}_{p=1}\frac{1}{p!}
\left[\left(\frac{\partial^{p}T_{-}}{\partial r^{p}_{H}}\right)_{Q_{KR}=Q_{C},~r_{H}=r_{j}}\right]\left(r_{H}-r_{j}\right)^{p}
   \end{array}\ee
 Now making use of 
equation(\ref{cr2}) and equation(\ref{cr9a}) we get:
\be\begin{array}{llll}\label{cr16}
    \displaystyle \Phi_{KR}\left(r_{H}\right)-\Phi_{KR}\left(r_{j}\right)=\sum^{\infty}_{p=1}\frac{(-1)^{p}(p+1)Q_{KR}}{r^{2}_{j}{\bf X}^{\frac{p}{2}}_{2j}}
\frac{\left|\frac{T_{-}-T_{j-}}{T_{j-}}\right|^{\frac{p}{2}}}{\left(1+
\frac{1}{{\bf X}^{\frac{1}{2}}_{2j}}\left|\frac{T_{-}-T_{j-}}{T_{j-}}\right|^{\frac{1}{2}}\right)^{p+2}}
   \end{array}\ee
To maintain the inherent assumptions of equilibrium statistical mechanics here we truncate the series at $p=1$.
Consequently we have:
\be\begin{array}{llll}\label{cr17}
    \displaystyle \Phi_{KR}\left(r_{H}\right)-\Phi_{KR}\left(r_{j}\right)\sim
J_{j}\left|T_{-}-T_{j-}\right|^{\frac{1}{2}}
   \end{array}\ee
where $J_{j}$ is a Gauss-Bonnet coupling ($\alpha_{(5)}$) dependent constant. Comparing equation(\ref{cr14}) and equation(\ref{cr17}) the second
critical exponent associated with the singularity in the Kalb Ramond potential turns out to be $\beta=\frac{1}{2}$.

\subsection{\bf Critical exponent $\gamma$}
Next we want to determine the critical exponent $\gamma$ which is associated with the isothermal compressibility 
near the critical point $r_{H}=r_{j}$ at constant charge $Q=Q_{C}$ can be expressed as,
\be\begin{array}{llll}\label{cr18}
    \displaystyle K^{-1}_{T}\propto |T-T_{j}|^{-\gamma}
   \end{array}\ee
appearing in the context of equilibrium statistical mechanics. 
 Now making use of 
equation(\ref{cr2}) and equation(\ref{cr9a}) in the context of Kalb Ramond black hole we get:
\be\begin{array}{llll}\label{cr19}
    \displaystyle K^{-1}_{T_{-}}=\frac{N_{j}T^{\frac{1}{2}}_{j-}}{\left|T_{-}-T_{j-}\right|^{\frac{1}{2}}}
   \end{array}\ee
where $N_{j}$ is a Gauss-Bonnet coupling ($\alpha_{(5)}$) dependent constant. Comparing equation(\ref{cr18}) and equation(\ref{cr19}) the third
critical exponent associated with the singularity in the isothermal compressibility calculated from Kalb Ramond antisymmetric tensor field turns out to be $\gamma=\frac{1}{2}$.


\subsection{\bf Critical exponent $\delta$}

In this subsection we evaluate the critical exponent ($\delta$) which is associated  with the electric potential $(\Phi)$
for the fixed value of the temperature at the  critical point $T = T_{j}$ as,
\be\begin{array}{llll}\label{bht1}
    \displaystyle \Phi(r_{H})-\Phi(r_{j})\sim |Q-Q_{j}|^{\frac{1}{\delta}}
   \end{array}\ee
where $Q$ be the corresponding electric charge. Here $Q_{j}$ is the associated charge evaluated at $r=r_{j}$.
 To evaluate $\delta$ we expand the Kalb Ramond charge $Q_{KR}(r_{H})$ in a
sufficiently small neighborhood of $r_{H} = r_{j}$ which gives,
\be\begin{array}{llll}\label{bht2}
    \displaystyle Q_{KR}(r_{H})=\sum^{\infty}_{q=0}\frac{1}{q!}
\left[\left(\frac{\partial^{q}Q_{KR}}{\partial r^{q}_{H}}\right)_{T_{-}=T_{j-},~r_{H}=r_{j}}\right]\left(r_{H}-r_{j}\right)^{q}
   \end{array}\ee
Now making use of the implicit functional dependence of the temperature
$T_{-}=T_{-}(r_{H},Q_{KR})$ we get the following constraint on the first derivative appearing in equation(\ref{bht2}) given by 
\be\begin{array}{llll}\label{bht3}
    \displaystyle\left(\frac{\partial Q_{KR}}{\partial r_{H}}\right)_{T_{-}=T_{j-},~r_{H}=r_{j}}
=-\left[\left(\frac{\partial T_{-}}{\partial r_{H}}\right)_{Q_{KR}}\right]_{r_{H}=r_{j}}
\left(\frac{\partial Q_{KR}}{\partial T_{-}}\right)_{r_{H}=r_{j}}=0.
   \end{array}\ee
Following similar prescription mentioned in equation(\ref{cr2}) in this context in the vicinity of critical point we can express the 
Kalb Ramond charge as
\be\begin{array}{llll}\label{bht4}
    \displaystyle Q_{KR}(r_{H})=Q_{KR}(r_{j})\left(1+Z_{C}\right)
   \end{array}\ee
with $Z_{C}<<1$. This results in
\be\begin{array}{llll}\label{bht5}
    \displaystyle Z^{\frac{1}{w}}_{C}=\sqrt[w]{\sum^{\infty}_{q=2}\frac{{\cal {\bf M}}_{jq}r^{q}_{j}\Delta^{q}_{C}}{q!Q_{jKR}}}
   \end{array}\ee
where $w$ be the order of truncation of the above mentioned Taylor series. Here 
${\cal {\bf M}}_{jq}=\left(\frac{\partial^{q}Q_{KR}}{\partial r^{q}_{H}}\right)_{T=T_{j},~r_{H}=r_{j}}$. Applying the additional constraints from 
equilibrium statistical mechanics we get $w=q=2$ in this context. This implies
\be\begin{array}{llll}\label{bht6}
    \displaystyle \Delta_{C}=\frac{2 \sqrt{Z_{C}}Q_{jKR}}{\sqrt{{\cal {\bf M}}_{j2}}r_{j}}.
   \end{array}\ee
Now in general Kalb Ramond potential is implicit function of Kalb Ramond charge and the corresponding horizon calculated from the metric function.
Consequently we have
\be\begin{array}{llll}\label{jki1}
    \displaystyle \left(\frac{\partial \Phi_{KR}}{\partial r_{H}}\right)_{T_{-}=T_{j-},~r_{H}=r_{j}}=
\left(\frac{\partial \Phi_{KR}}{\partial r_{H}}\right)_{Q_{KR}=Q_{jKR},~r_{H}=r_{j}}.
   \end{array}\ee
Next expanding the Kalb Ramond potential in Taylor series in the neighborhood of $r_{H}=r_{j}$ we get:
\be\begin{array}{llll}\label{bht71}
    \displaystyle \Phi_{KR}\left(r_{H}\right)-\Phi_{KR}\left(r_{j}\right)=\sum^{\infty}_{s=1}
\left[\left(\frac{\partial^{s}\Phi_{KR}}{\partial r^{s}_{H}}\right)_{T_{-}=T_{j-},~r_{H}=r_{j}}\right]\left(r_{H}-r_{j}\right)^{s}\\
\displaystyle~~~~~~~~~~~~~~~~~~~~~~~~~~~~=\sum^{\infty}_{s=1}\frac{(-1)^{s}(s+1)Q_{C}\left(2!\right)^{s}Q^{s}_{Cj}Z^{\frac{s}{2}}_{C}}
{r^{s+2}_{j}{\bf M}^{\frac{s}{2}}_{j2}\left(1+\frac{2!\sqrt{Z_{C}}Q_{Cj}}{\sqrt{{\bf M}_{j2}}r_{j}}\right)^{s+2}}.
   \end{array}\ee
Applying the constraints from equilibrium statistical mechanics we truncate the Taylor series at $s=1$. This implies
\be\begin{array}{llll}\label{bht72}
    \displaystyle \Phi_{KR}\left(r_{H}\right)-\Phi_{KR}\left(r_{j}\right)
=\frac{(-4)Q_{C}Q_{Cj}\sqrt{Z_{C}}}
{r^{3}_{j}\sqrt{{\bf M}_{j2}}\left(1+\frac{2!\sqrt{Z_{C}}Q_{Cj}}{\sqrt{{\bf M}_{j2}}r_{j}}\right)^{3}}.
   \end{array}\ee
Substituting equation(\ref{bht6}) in equation(\ref{bht71}) and taking upto the leading order contribution we get:
\be\begin{array}{llll}\label{bht73}
    \displaystyle \Phi_{KR}\left(r_{H}\right)-\Phi_{KR}\left(r_{j}\right)
=\frac{(-4)Q_{C}Q_{Cj}\left|Q_{KR}-Q_{jKR}\right|^{\frac{1}{2}}}
{r^{3}_{j}\sqrt{Q_{jKR}{\bf M}_{j2}}\left(1+\frac{2!\left|Q_{KR}-Q_{jKR}\right|^{\frac{1}{2}}Q_{Cj}}{\sqrt{Q_{jKR}{\bf M}_{j2}}r_{j}}\right)^{3}}\approx B_{j}\left|Q_{KR}-Q_{jKR}\right|^{\frac{1}{2}}
   \end{array}\ee
where $B_{j}=\frac{(-4)Q_{C}Q_{Cj}}{r^{3}_{j}\sqrt{Q_{jKR}{\bf M}_{j2}}}$. Equating equation(\ref{bht73}) with equation(\ref{bht1}) the corresponding
critical exponent on the Kalb Ramond potential turns out to be $\delta=2$.

\subsection{\bf Critical exponent $\varphi$}

Next our aim is to evaluate the critical exponent $\varphi$ from the relationship between specific heat at constant charge ($C_{Q}$) and
electric charge of the black hole given by:
\be\begin{array}{llll}\label{mcg1}
    \displaystyle C_{Q}\sim \frac{1}{\left|Q-Q_{j}\right|^{\varphi}}.
   \end{array}\ee
For the Kalb Ramond black hole using equation(\ref{cr12}) we get
\be\begin{array}{llll}\label{cht1}
    \displaystyle C^{-}_{Q_{KR}}
=\left[\frac{D_{j}}{\Delta_{C}\sqrt{{\bf X}_{2j}}}\right]=\frac{V_{j}}{\left|Q_{KR}-Q_{jKR}\right|^{\frac{1}{2}}}
   \end{array}\ee
 where $V_{j}=\frac{D_{j}\sqrt{{\cal {\bf M}}_{j2}}r_{j}}{2\sqrt{Q_{jKR}{\bf X}_{2j}}}$. Comparing
equation(\ref{mcg1}) and equation(\ref{cht1}) the critical exponent turns out to be $\varphi=\frac{1}{2}$.

\subsection{\bf Critical exponent $\psi$}
Further we want to determine the critical exponent $\psi$ which is associated to the entropy ($S$) as given by,
\be\begin{array}{llll}\label{cr14v}
    \displaystyle S\left(r_{H}\right)-S\left(r_{j}\right)\propto |Q-Q_{j}|^{\psi}
   \end{array}\ee
appearing in the context of equilibrium statistical mechanics. To serve this purpose
 we Taylor expand the Kalb Ramond entropy in the neighborhood of the critical point $r_{H}=r_{j}$ yields,
\be\begin{array}{llll}\label{cr15v}
    \displaystyle S_{KR}\left(r_{H}\right)-S_{KR}\left(r_{j}\right)=\sum^{\infty}_{t=1}\frac{1}{t!}
\left[\left(\frac{\partial^{t}S_{H}}{\partial r^{t}_{H}}\right)_{Q_{KR}=Q_{C},~r_{H}=r_{j}}\right]\left(r_{H}-r_{j}\right)^{t}
   \end{array}\ee
 Now by making use of 
equation(\ref{bht6}) we get:
\be\begin{array}{llll}\label{cr16v}
    \displaystyle S_{KR}\left(r_{H}\right)-S_{KR}\left(r_{j}\right)=
\sum^{\infty}_{t=1}\frac{3\pi^2 }{t!}\left(\frac{{\bf A}_{t}r^{3}_{j}}{2^{2-t}}+{\bf B}_{t}\alpha^{2-t}_{(5)}r^{t}_{j}\right) \left(\frac{2\sqrt{Z_{C}}Q_{jKR}}{\sqrt{{\bf M}_{j2}}}\right)^{t}
\left(1+
\frac{2Q_{jKR}\sqrt{Z_{C}}}{\sqrt{{\bf M}_{j2}}r_{j}}\right)^{3-t}\\
\displaystyle~~~~~~~~~~~~~~~~~~~~~~~~~~~=\sum^{\infty}_{t=1}\frac{3\pi^2 }{t!}\left(\frac{{\bf A}_{t}r^{3}_{j}}{2^{2-t}}+{\bf B}_{t}\alpha^{2-t}_{(5)}r^{t}_{j}\right) 
\left(\frac{2\sqrt{\left|Q_{KR}-Q_{jKR}\right|
Q_{jKR}}}{\sqrt{{\bf M}_{j2}}}\right)^{t}
\left(1+
\frac{2\sqrt{Q_{jKR}\left|Q_{KR}-Q_{jKR}\right|}}{\sqrt{{\bf M}_{j2}}r_{j}}\right)^{3-t}
   \end{array}\ee
where the expansion co-efficients $({\bf A}_{t},{\bf B}_{t})\forall t$ are givan by:
\be\label{ghtaa4}
{\bf A}_{t}=
\left\{
	\begin{array}{ll}
                    \displaystyle 1 & \mbox{ :{\bf t=1,2}}  \\
         \displaystyle  0 & \mbox{~:{\bf t$>$2 }} 
          \end{array}
\right.
\ee
and 
\be\label{ghtaa4q}
{\bf B}_{t}=
\left\{
	\begin{array}{ll}
                    \displaystyle 1 & \mbox{ :{\bf t=1}}  \\
         \displaystyle  0 & \mbox{~:{\bf t$\geq$2 }} .
          \end{array}
\right.
\ee
To maintain the inherent assumptions of equilibrium statistical mechanics here we truncate the series at $t=1$.
Taking upto leading order contribution in equation(\ref{cr16v}) we get:
\be\begin{array}{llll}\label{cr17v}
    \displaystyle S_{KR}\left(r_{H}\right)-S_{KR}\left(r_{j}\right)\approx
L_{j}\left|Q_{KR}-Q_{jKR}\right|^{\frac{1}{2}}
   \end{array}\ee
where $L_{j}=\frac{6\pi^2\sqrt{Q_{jKR}}}{\sqrt{{\bf M}_{j2}}}\left(\frac{r^{3}_{j}}{2}+\alpha_{(5)}r_{j}\right)$. Comparing equation(\ref{cr14v}) and equation(\ref{cr17v}) the second
critical exponent associated with the singularity in the Kalb Ramond potential turns out to be $\psi=\frac{1}{2}$.
\subsection{\bf Critical exponent $\nu$ and $\eta$}

At last we would like to determine the rest of the two critical exponents $\nu$ and $\eta$ which
are associated with the correlation length ($\zeta$) and two point correlation function respectively. Now keeping 
the basic assumptions of equilibrium statistical mechanics intact we can write:

\be\begin{array}{lll}\label{xilo1}
    \displaystyle \gamma=\nu(2-\eta),~~~(2-\alpha)=\nu d
   \end{array}\ee

in our setup. Here d is spatial dimensionality of our concerned setup. For our problem $d=3$ and substituting this into equation(\ref{xilo1}) we find
$\nu=\frac{1}{2}$ and $\eta=1$. Since equation(\ref{xilo1}) is spatial dimension dependent relation 
it may happen that in $d>3$ or $d\leq 2$ the critical exponents estimated from 
Gauss-Bonnet gravity in presence of charged Kalb Ramond tensor field is completely different from $d=3$ value.


\subsection{\bf Thermodynamic Scaling Laws}
We have explicitly calculate the values of the critical exponents $\alpha,\beta,\gamma,\delta,\varphi$ and $\psi$ associated
with the discontinuities of various thermodynamic variables. Sets of all critical exponents satisfies the following 
mathematical consistency conditions:
\be\begin{array}{llll}\label{vv1}
    \displaystyle \alpha+2\beta+\gamma=2,\\
\displaystyle \alpha+\beta(\delta+1)=2,\\
\displaystyle (2-\alpha)(\delta\psi-1)+1=(1-\alpha)\delta,\\
\displaystyle \gamma(\delta+1)=(2-\alpha)(\delta-1),\\
\displaystyle \gamma=\beta(\delta-1),\\
\displaystyle \delta(\varphi+2\psi-1)=1\\
   \end{array}\ee
in the context of charged Kalb Ramond black hole thermodynamics. These conditions are exactly synonymous version of 
the {\it Rushbrooke-Josephson scaling laws}\cite{golden,ma,hohen} in the context of equilibrium statistical mechanics. Now keeping the appearance of spatial dimensionality
in equation(\ref{xilo1}) we can say that the above mentioned scaling laws cannot holds good for $d\leq 2$ and $d>3$. So that 
the {\it Rushbrooke-Josephson scaling laws}\cite{golden,ma,hohen} are not at all universal in all spatial dimension. Universality are maintained only at $d=3$.

\section{\bf Widom scaling via Generalized Homogeneous Function Hypothesis Test}

According to the statement of the `` Generalized Homogeneous Function Hypothesis Test'' (GHFHT)\cite{lusto,Wu,stanley,diba5}
for the black holes in the neighborhood of the critical point, the singular
part of the thermodynamic free energy is a generalized homogeneous function of
its characteristic variables. Let we start with 
the well known free energies from which our aim is to calculate the degree of homogeneity
from the singular part of the free energy. This implies
\be\begin{array}{llll}\label{cvcv1}
    \displaystyle F_{KR}(\varGamma^{a}T_{-},\varGamma^{b}Q_{KR})=\varGamma F_{KR}(T_{-},Q_{KR})~\Rightarrow ~F_{KR}(\varGamma^{a}\epsilon_{C},\varGamma^{b}Z_{C})
=\varGamma F_{KR}(\epsilon_{C},Z_{C}),\\
    \displaystyle G_{KR}(\chi^{c}T_{-},\chi^{d}\Phi_{KR})=\chi G_{KR}(T_{-},\Phi_{KR})~\Rightarrow ~G_{KR}(\chi^{c}\epsilon_{C},\chi^{d}Z_{C})
=\chi G_{KR}(\epsilon_{C},Z_{C}),\\
    \displaystyle H_{KR}(\vartheta^{e}S_{H},\vartheta^{f}\Phi_{KR})=\vartheta H_{KR}(S_{H},\Phi_{KR})~\Rightarrow ~H_{KR}(\vartheta^{g}\epsilon_{C})
=\vartheta H_{KR}(\epsilon_{C}),\\
    \displaystyle M_{KR}(\lambda^{h}S_{H},\lambda^{t}Q_{KR})=\lambda M_{KR}(S_{H},Q_{KR})~\Rightarrow ~M_{KR}(\lambda^{u}\epsilon_{C})
=\lambda M_{KR}(\epsilon_{C}).
   \end{array}\ee
Since $\epsilon_{C},Z_{C}<<1$ we can expand the above mentioned free energies in Taylor series in the neighborhood of critical point.
This gives
\be\begin{array}{llll}\label{vbvb1} 
    \displaystyle O_{KR}(x^{(1)},x^{(2)})=\sum^{\infty}_{m=0}\left\{\frac{1}{m!}
\left[\sum^{2}_{k=1}\left(x^{(k)}-x^{(k)}_{j}\right)\partial_{x^{'(k)}}\right]^{m}O_{KR}(x^{'(1)},x^{'(2)})\right\}_{x^{'(1)}=x^{(1)}_{j},x^{'(2)}=x^{(2)}_{j}}
   \end{array}
\ee
where 
\be\label{normala}
 O_{KR}(x^{(1)},x^{(2)})=
\left\{
	\begin{array}{ll}
                    \displaystyle F_{KR} & \mbox{ \it with $x^{(1)}=T_{-},~x^{(2)}=Q_{KR}$}  \\
         \displaystyle  G_{KR} & ~\mbox{\it with $x^{(1)}=T_{-},~x^{(2)}=\Phi_{KR}$}\\
\displaystyle H_{KR} & ~\mbox{\it with $x^{(1)}=S_{H},~x^{(2)}=\Phi_{KR}$}\\
\displaystyle M_{KR} & ~\mbox{\it with $x^{(1)}=S_{H},~x^{(2)}=Q_{KR}$}.
          \end{array}
\right.
\ee
Now collecting the singular contribution appearing in equation(\ref{vbvb1}) we get: 
\be\begin{array}{llll}\label{vbvb2} 
    \displaystyle O^{\infty}_{KR}(x^{(1)},x^{(2)})=\left\{\frac{1}{2}\sum^{2}_{k=1}
\left[\partial_{x^{'(k)}}O_{KR}(x^{'(1)},x^{'(2)})\right]^{2}\left(x^{(k)}-x^{(k)}_{j}\right)^{2}\right\}_{x^{'(1)}=x^{(1)}_{j},x^{'(2)}=x^{(2)}_{j}}
   \end{array}
\ee
where individually singular contribution reads
\be\label{normala1}
 O^{\infty}_{KR}(x^{(1)},x^{(2)})=
\left\{
	\begin{array}{ll}
                    \displaystyle F^{\infty}_{KR}={\cal A}^{(1)}_{j}|\epsilon_{C}|^{\frac{1}{a}}+{\cal B}^{(1)}_{j}Z^{\frac{1}{b}}_{C} & \mbox{ \it with $a=b=\frac{2}{3}$}  \\
         \displaystyle  G^{\infty}_{KR}={\cal A}^{(2)}_{j}|\epsilon_{C}|^{\frac{1}{c}}+{\cal B}^{(2)}_{j}Z^{\frac{1}{d}}_{C} & ~\mbox{\it with $c=d=\frac{2}{3}$}\\
\displaystyle H^{\infty}_{KR}={\cal A}^{(3)}_{j}|\epsilon_{C}|^{\frac{1}{e}}Z^{\frac{1}{v}}_{C}+{\cal B}^{(3)}_{j}Z^{\frac{1}{f}}_{C} & ~\mbox{\it with $e=2,~v=1,~f=\frac{2}{3}$}\\
\displaystyle M^{\infty}_{KR}={\cal A}^{(4)}_{j}|\epsilon_{C}|^{\frac{1}{g}}Z^{\frac{1}{h}}_{C} & ~\mbox{\it with $g=-2,~h=\frac{1}{2}$}.
          \end{array}
\right.
\ee
This directly shows the exponent appearing in $|\epsilon_{C}|$ and $Z_{C}$ are not exactly same always in the singular part of the free energy.
Most importantly Gibbs free energy and Helmholtz free energy are the exceptions where they behaves as a usual homogeneous functions.
 
The exponents appearing in the singular part of Gibbs free energy and Helmholtz free energy are connected with the critical exponents
evaluted for the charged Kalb Ramond black hole as:
\be\begin{array}{llll}\label{criop1}
    \displaystyle a(2-\alpha)=1,~ c(2-\alpha)=1,\\
\displaystyle \beta a + b=1,~ \beta c +d=1,\\
\displaystyle \delta (1-b)=b,~\delta (1-d)=d,\\
\displaystyle \gamma a +1=2b,~\gamma c +1=2d,\\
\displaystyle \psi b+a=1,~\psi d+c=1,\\
\displaystyle \varphi b+1=2a,~\varphi d+1=2c
   \end{array}\ee
which are usual consistency relations satisfied in the context of equilibrium statistical mechanics. 
These are commonly known as {\it Widom scaling} hypothesis. On the other hand for Enthalpy and Black hole mass
the above mentioned consistency conditions are modified as:
\be\begin{array}{llll}\label{criop2}
    \displaystyle e(2-\alpha)=3v,~ g(2-\alpha)=-3,\\
\displaystyle \beta e + \frac{1}{f}=\frac{5v}{2},~ \beta g +h=-\frac{1}{2},\\
\displaystyle \delta \left(1-\frac{1}{f}\right)=-v,~\delta (1-h)=-hg,\\
\displaystyle \gamma e +2v=\frac{2}{f},~\gamma g +2=2h,\\
\displaystyle \psi f+e=\frac{7v}{3},~\frac{\psi}{h}+g=-1,\\
\displaystyle \varphi f+v=\frac{2e}{3},~\frac{\varphi}{h} +1=-g
 \end{array}\ee
which are obviously a new results in the context of charged Kalb Ramond black hole induced by the 5D Gauss-Bonnet coupling ($\alpha_{(5)}$).

\section{AdS/CMT realization in presence of five dimensional Gauss-Bonnet coupling}
The AdS/CFT correspondence \cite{malda2,malda5,klebanov}
has yielded many important insights into 
the present research on the several branches of field theory. Among various results obtained so far, one of the most significant is the
universality of the ratio of the shear viscosity ($\eta$) to the entropy density ($s$) given by \cite{poly,kovtan,buchel,son}:
\be\begin{array}{lll}\label{shr1}
    \displaystyle \frac{\eta}{s}=\frac{1}{4\pi}
   \end{array}\ee
for strongly coupled gauge theories with an Einstein gravity (holographic) dual in the limit $N\rightarrow\infty$ and $\lambda\rightarrow\infty$ (large N limit).
 In this context N represents the number of
colors and $\lambda$ be the well known 't Hooft coupling.  
It was further conjectured that equation(\ref{shr1}) hass a universal lower bound, commonly known as the
Kovtun-Starinets-Son (KSS) bound satisfied by all known substances including water and liquid helium
so far. Most importantly it also includes the quark-gluon plasma created at Relativistic
Heavy Ion Collider (RHIC) \cite{tean,song,dusling,adare} and certain cold atomic gases in the unitarity limit. 
For pure gluon inspired QCD numerical value for the $\frac{\eta}{s}$ is 0.3 above the deconfinement temperature \cite{hbmayer,cohen,son1,cher,foux,buchel1,beni}.
 More generally, string theory 
contains higher derivative perturbative terms in the action which is 
obtained from string loop corrections (via CFT disk amplitudes), inclusion of which will modify the ratio. In
terms of gauge theories, such modifications are appearing via $\frac{1}{\lambda}$ or $\frac{1}{N}$ corrections. 
 So far it was found
that the correction is consistent with the conjectured bound.

In this section, instead of knowing specific string theory corrections, we explore the modification of
equation(\ref{shr1}) due to generic five dimensional Gauss-Bonnet term in presence of antisymmetric Kalb Ramond tensor field 
in the holographic gravity dual. String two loop corrections can also generate such terms, but
they are suppressed by powers of $g_{s}$.
 Specifically in our model in presence of Gauss-Bonnet coupling ($\alpha_{(5)}$) equation(\ref{shr1}) is modified as \cite{myers}
\be\begin{array}{llll}\label{shr2}
    \displaystyle \left(\frac{\eta}{s}\right)_{KR}=\frac{1}{4\pi}\left(1-4\alpha_{(5)}\right)+{\cal O}(\alpha^{2}_{(5)})
   \end{array}\ee
which is independent of the magnitude of the Kalb Ramond charge since the Gauss-Bonnet gravity sector is non-interacting with the 
rank 3 Kalb Ramond antisymmetric tensor fields. If we allow such non-trivial higher order interactions 
then equation(\ref{shr2}) involves two extra correction terms in the $\frac{\eta}{s}$ ratio. One of them 
is the product of Kalb Ramond charge $Q_{KR}$ and the five dimensional Gauss-Bonnet coupling $\alpha_{(5)}$
originated through interaction between gravitons and Kalb Ramond fields (Diffeomorphism invariant interaction$\rightarrow {\cal C}_{KR}{\cal L}_{GB}{\cal H}_{ABC}{\cal H}^{ABC}$
where the mass dimension of the coupling ${\cal C}_{KR}$ is $M^{-2}_{PL}$). On the other side there is another term 
coming from the self interactions (quadratic interaction$\rightarrow {\cal H}_{ABC}{\cal H}^{ABC}$)
 between Kalb Ramond fields which plays a crucial role in the present context.
Consequently equation(\ref{shr2}) is modified as: 
\be\begin{array}{llll}\label{shr2cv}
    \displaystyle \left(\frac{\eta}{s}\right)_{KR}=\frac{1}{4\pi}\left(1-4\alpha_{(5)}+4\delta\alpha_{(5)}\right)+{\cal O}(\alpha^{2}_{(5)},
Q^{2}_{KR}M^{4}_{PL}).
   \end{array}\ee
where where $\delta=\frac{{\cal C}_{KR}Q_{KR}M^{4}_{PL}}{4}$ is the correction factor appearing due to interaction between graviton and Kalb Ramond fields. Without any interactions
$\delta$ become zero.
 Both the equation(\ref{shr2}) and equation(\ref{shr2cv}) show slight deviation from its $\alpha_{(5)}\rightarrow 0$ result. These results are 
the manifestation of {\it Kubo formula} in AdS/CFT \cite{myers}. In particular, the viscosity bound is strictly violated as
 $\alpha_{(5)}\rightarrow\frac{1}{4(1-\delta)}$ where the total off-shell action becomes zero. It is likely the on-shell action also vanishes, implying that the correlation
function calculated from the energy momentum tensor in the boundary 4D dual CFT theory could become identically zero in this limit.
Most importantly, for $\alpha_{(5)}<\frac{1}{4(1-\delta)}$ the background five dimensional Gauss-Bonnet gravity has a boundary 
dual CFT and $\frac{\eta}{s}$ cannot be negative. 
Here $\frac{\eta}{s}\rightarrow 0$ is only allowed when bulk causality or unitarity is violated which is not true for our
 setup. The correction appearing in equation(\ref{shr2}) is only significant in five dimension. If the dimensionality of the space-time is lower than 5 then 
Gauss-Bonnet term is topologically invariant and consequently no such corrections will contribute in the $\frac{\eta}{s}$ ratio. For higher dimensional 
Lovelock gravity, which is the generalized version of Gauss-Bonnet gravity in $D>5$, one can see that the causality is not enough to prevent
this behavior for $D>10$ \cite{jose1}. An interesting situation 
like  $\frac{\eta}{s}\rightarrow 0$  can be realized when $D$ goes to infinity \cite{jose2}.

The shear viscosity of the boundary CFT is associated with absorption of transverse modes by
the black brane in the bulk dominated by antisymmetric Kalb Ramond tensor fields. This is a natural picture since the shear viscosity measures the dissipation rate of
the quantum fluctuations and predicts the fact that the faster absorption in the black brane leads to the higher the dissipation rate. For example,
as $\alpha_{(5)}\rightarrow -\infty$, the ratio $\frac{\eta}{s}\rightarrow \infty$ which describes a situation where every bit of the black brane horizon
devours the transverse fluctuations very quickly. In this limit the curvature singularity 
approaches the horizon and the tidal force near the horizon becomes very very strong. On the contrary, as 
$\alpha_{(5)}\rightarrow\frac{1}{4(1-\delta)}$
the black brane very slowly absorbs transverse modes. Most importantly, at exactly  
 $\alpha_{(5)}=\frac{1}{4(1-\delta)}$
the radial direction of the background geometry resembles a Ba$\tilde{n}$ados-
Teitelboim-Zanelli (BTZ) black brane \cite{myers1,lee1,storm,esko,myung}. 
Now from the figure(\ref{fig3}) it is obvious that to get
 a positive definite numerical value of the Hawking temperature we have to fix the lower bound
of the the five dimensional Gauss-Bonnet parameter $\alpha_{(5)}$. The lower bound or cutoff for 
$\alpha_{(5)}$ is estimated as $\alpha^{C}_{(5)}=4.6\times 10^{-6}$. On the
other hand the upper bound of $\alpha_{(5)}$ is always less than $\frac{1}{4(1-\delta)}$. Combining all these physical situations we can write:
\be\begin{array}{llll}\label{vutc}
    \displaystyle \alpha^{C}_{(5)}\leq \alpha_{(5)}< \frac{1}{4\left(1-\delta\right)}.
   \end{array}\ee

  Now to maintain the crucial constarint appearing from third law of thermodynamics in the context of charged Kalb Ramond black hole
the equality in the equation(\ref{vutc}) is relaxed. Consequently the physical bound for the five dimensional Gauss-Bonnet coupling 
turns out to be:
\be\begin{array}{llll}\label{vutc1}
    \displaystyle \alpha^{C}_{(5)}< \alpha_{(5)}< \frac{1}{4\left(1-\delta\right)}.
   \end{array}\ee
Considering equation(\ref{vutc1}), equation(\ref{shr2}) and equation(\ref{shr2cv}) the physical bound on 
$\frac{\eta}{s}$ ratio for charged Kalb Ramond black hole in presence of five dimensional Gauss-Bonnet gravity
is given by:
\be\begin{array}{lll}\label{etas}
    \displaystyle    \left(\frac{\eta}{s}\right)^{C}_{KR} >\left(\frac{\eta}{s}\right)_{KR}>0
   \end{array}\ee
where 
 $\left(\frac{\eta}{s}\right)^{C}_{KR}=[
 0.079576007$ (with $\delta=0$), 
 $0.079576373$ (with $\delta=\frac{1}{4}$)]. Here for $\delta=\frac{1}{4}$ we use ${\cal C}_{KR}Q_{KR}=1$ in Planckian unit. Now fixing $Q_{KR}$ at the lower cut-off charge $Q^{C}_{KR}$ the corresponding numerical 
value of the coupling ${\cal C}_{KR}$ turns out to be $1.03\times 10^{4}$.
But direct/indirect signatures of such strong coupling not yet detected at LHC. So it is better for the
 phenomenological purpose to relax the condition ${\cal C}_{KR}Q_{KR}=1$. The perturbative analysis suggests that feasible phenomenological features 
may be observe at future run of LHC \cite{lhc} or any future linear collider (ILC) \cite{ilc} provided ${\cal C}_{KR}Q_{KR}<<1$ and $\delta<<\frac{1}{4}$. Including the constraint from the lower cut-off charge 
at $Q^{C}_{KR}$ the corresponding numerical bound on the coupling translates into ${\cal C}_{KR}<<1$ in the weak coupling regime. 
This implies that in equation(\ref{etas})  $\left(\frac{\eta}{s}\right)^{C}_{KR}=
 0.079576026$ (with $\delta=1.8\times10^{-10}$).
On the other hand without Gauss-Bonnet gravity using equation(\ref{shr1}) the $\frac{\eta}{s}$
ratio is given by 
$\left(\frac{\eta}{s}\right)_{GR}=0.079577471$.
This implies that the fractional deviation with respect to the lower bound of $\frac{\eta}{s}$ can be estimated as:
\be\label{nox2}
{\bf\Delta^{C}_{GR}}= \left[\frac{ \left(\frac{\eta}{s}\right)_{GR}- \left(\frac{\eta}{s}\right)^{C}_{KR}}{ \left(\frac{\eta}{s}\right)_{GR}}\right]=
\left\{
	\begin{array}{ll}
                    \displaystyle 1.839\times 10^{-5} & \mbox{ \it with $\delta=0$}  \\
         \displaystyle 1.379\times 10^{-5} & ~\mbox{\it  with $\delta=0.25$}\\
\displaystyle 1.815\times 10^{-5} & ~\mbox{\it  with $\delta=1.8\times10^{-10}$}.
          \end{array}
\right.
\ee
Now at the horizon $r_{H}=0.038~M^{-1}_{PL}$ we have $\alpha_{(5)}=0.00095$ (as shown in $r_{H}$ vs $\alpha_{(5)}$ plot)
 the corresponding $\frac{\eta}{s}$ ratio is given by $\left(\frac{\eta}{s}\right)_{KR}=[0.079275077~~(with~~\delta=0),0.079350675~~(with~~\delta=0.25),
0.079275109~~(with~~\delta=1.8\times10^{-10})]$.
 Consequently the
fractional deviation in the $\frac{\eta}{s}$ ratio is measured as:
\be\label{nox3}
    \displaystyle {\bf\Delta_{GR}}=
 \left[\frac{ \left(\frac{\eta}{s}\right)_{GR}- \left(\frac{\eta}{s}\right)_{KR}}{ \left(\frac{\eta}{s}\right)_{GR}}\right]
=\left\{
	\begin{array}{ll}
                    \displaystyle 3.799\times 10^{-3} & \mbox{ \it with $\delta=0$}  \\
         \displaystyle 2.850\times 10^{-3} & ~\mbox{\it  with $\delta=0.25$}\\
\displaystyle 3.797\times 10^{-3} & ~\mbox{\it  with $\delta=1.8\times10^{-10}$}.
          \end{array}
\right.\ee
This implies at the horizon $r_{H}=0.038~M^{-1}_{PL}$, ${\bf \Delta_{GR}}>>{\bf\Delta^{C}_{GR}}$
i.e. the fractional deviation with respect to the GR limiting result is very large compared 
to the deviation for the lower cutoff. Additionally here we have to mention that at the point of phase transition the 
critical value of the five dimensional Gauss-Bonnet coupling $\alpha^{Crit}_{(5)}=4.13\times 10^{-5}\neq\alpha^{C}_{(5)}$ 
and the consequently at the critical point we have 
$\left(\frac{\eta}{s}\right)^{Crit}_{KR}=[0.079564325(with~\delta=0),0.079567611(with~\delta=0.25),0.079564339(with~\delta=1.8\times10^{-10})$].

\section{Summary and outlook}
In this article we have made a comprehensive study of AdS black hole thermodynamics and equilibrium statistical mechanics 
inspired from string theory motivated rank 3 antisymmetric Kalb Ramond tensor field
 and its implications on phase transition and AdS/CFT correspondence. 
Our model includes a perturbation of the Einstein gravity characterized by the presence of quadratic correction
 appearing as Gauss-Bonnet coupling in five dimension which also includes the effect of string two loop 
correction in the gravity sector. Our study centered around three distinct aspects :-\\
\begin{itemize}
\item Determining the physically acceptable metric function ($h_{-}(r)$), its asymptotic behaviour and study of branch singularity and killing horizon from $h_{-}(r)$.\\
\item Study of different thermodynamic parameters to examine their behaviour in presence of charged Kalb Ramond field and Gauss-Bonnet coupling ($\alpha_{(5)}$) both
in the context of black hole thermodynamics.\\
\item Estimating  the values of the critical exponents associated with the discontinuities
in various thermodynamic parameters and the corresponding order of the phase transition in AdS space-time.\\ 
\item Comments on the validity of {\it Rushbrooke Josephson scaling laws} in the present setup.\\
\item Study of Widom scaling hypothesis via GHFHT by establishing the connection between the degree of homogeneity in free energy with the critical exponents.\\
\item Establishing  the relation between five dimensional Gauss-Bonnet coupling ($\alpha_{(5)}$) with the well known $\frac{\eta}{s}$ ratio
appearing in the 4D CFT holographic dual theory. \\
\item Determining the physical bound on  $\frac{\eta}{s}$ ratio by fixing the lower cutoff and upper cutoff of $\alpha_{(5)}$ from the 
thermodynamical behaviour in the bulk theory.
\end{itemize} 

Our results can be summarized as follows :\\
\begin{itemize}
\item Positive branch solution of the metric function ($h_{+}(r)$) produces naked singularity which violates the well known cosmic censorship.
To avoid this situation we further use the negative branch solution of the metric function ($h_{-}(r)$)
from which we determine the branch singularity and killing horizon by making use of all possible signatures of 
five dimensional cosmological constant $\Lambda_{(5)}$ and positive signature of ADM mass parameter $\mu$ for
different values of the Kalb Ramond charge $Q_{KR}$. \\

\item We have explicitly shown the behaviour of killing horizon ($r_{H}$) with respect to the five dimension Gauss-Bonnet coupling for $\Lambda_{(5)}<0$.
We have shown in a plot that as $\alpha_{(5)}$ increases the corresponding value of the killing horizon decreases
which is obviously an intersting feature in the present context.\\

\item We have determined the analytical expressions for various thermodynamic parameters i.e. Hawking temperature, Bekenstein Hawking entropy,
specific heat at constant Kalb Ramond charge, specific heat at constant Kalb Ramond potential, isothermal compressibility and volume expansivity. 

\item We have determined the lower cutoff of the five dimensional Gauss-Bonnet coupling by applying the constraint from the third law of thermodynamics on
the Hawking temperature. The numerical values of the lower bound of $\alpha_{(5)}$ is different for the different signatures of the five dimensional 
cosmological constant $\Lambda_{(5)}$.

\item We then explore the possibility of phase transition appearing in specific heat at constant $Q_{KR}$ for the -ve
 signatures of the five dimensional 
cosmological constant ($\Lambda_{(5)}<0$) in the variation with respect to $\alpha_{(5)}$. Most significantly the discontinuity
appearing in the $C^{-}_{Q_{KR}}$ vs $\alpha_{(5)}$ plot at the $\alpha^{Crit}_{(5)}\sim 4.13\times 10^{-5}$ (critical point).
The corresponding value of killing horizon for the fixed ADM mass parameter and Kalb Ramond charge turns out to be $r_{j}\sim 0.06~M^{-1}_{PL}$
which is frequently appearing in the analysis of critical exponents.
 We have also shown that for $\alpha_{(5)}<4.13\times 10^{-5}$ the corresponding specific heat approaches to negative value.

\item Next we have studied the possibility of phase transition appearing in $C^{-}_{\Phi_{KR}}$ vs $\alpha_{(5)}$ and $K^{-}_{T_{-}}$ vs $\alpha_{(5)}$ plots
 in AdS space-time.
 The specific heat $C^{-}_{\Phi_{KR}}$ approaches the negative value in the interval $0.000038<\alpha_{(5)}<0.000055$. Rest of the features are exactly 
same as that appearing in  $C^{-}_{Q_{KR}}$ vs $\alpha_{(5)}$ plot.
On the other hand for $\alpha_{(5)}<0.00018$, the inverse of the isothermal compressibility shows exactly the opposite behaviour as depicted in the variation of 
$C^{-}_{Q_{KR}}$.

\item Further we apply Legendre transformation technique to convert one free energy to the other. Such transformation rule is exactly similar
to that appearing in the context of equilibrium statistical mechanics. 
 
\item We have explicitly verified the applicability of Ehrenfest's theorem and determine the order of phase transition by making use of 
PD ratio in our setup. Corresponding order of the phase transition turns out to be 'two' which is obviously an important result for charged Kalb Ramond black hole
in the Gauss-Bonnet gravity. 

\item We then extend this idea to the theory of critical phenomena  by determining the critical exponents connected with the 
discontinuities of various thermodynamic parameters appearing in our setup. From the elaborate study of these parameters in
 the neighborhood of the critical point and maintaining the basic assumptions of equilibrium statistical mechanics, the critical exponents 
in our set up is calculated as:$\alpha=\frac{1}{2},\beta=\frac{1}{2},\gamma=\frac{1}{2},\delta=2,\varphi=\frac{1}{2},\psi=\frac{1}{2},\nu=\frac{1}{2}$ and $\eta=1$
where the spatial dimensionality is $d=3$. Most importantly all of these critical exponents satisfies the {\it Rushbrooke Josephson scaling laws} at $d=3$. But
for other than
$d=3$ we cannot appropriately comment on the universality of these scaling laws. It may happen that in some exceptional situations of the higher derivative gravity
theories for $d\neq 3$ universality is maintained.   

\item Finally, we give an upper bound on the $\alpha_{(5)}$ which is always less than $\frac{1}{4(1-\delta)}$ from the {\it Kubo formula}. Making use of the upper as well as the lower
bound of $\alpha_{(5)}$ we determine the bound on the $\frac{\eta}{s}$ ratio in the context of charged Kalb Ramond black hole in presence of Gauss-Bonnet gravity.
We also measure the fractional deviation from the value of $\frac{\eta}{s}$ ratio obtained from the Einstein gravity. Very near to the lower cutoff (just slightly above)
the amount of deviation from its value predicted from Einstein gravity is very very small. On the other hand in some intermediate value of $\alpha_{(5)}$ (let us say at the outer horizon)
the amount of deviation is larger compared to the previous one. Most importantly, at the critical point the numerical value of $\frac{\eta}{s}$ exactly converges towards 
the lower cutoff of $\alpha_{(5)}$.
\end{itemize}
   
Some interesting open issues in this context of the present study can be  
similar  investigations on various aspects of the four dimensional holographic CFT dual when the bulk contains antisymmetric tensor field strength  
of rank higher than three. Such fields are quite generic in string theoretic models and we propose to study them in some future work.


\section*{Acknowledgments}

SC thanks Council of Scientific and
Industrial Research, India for financial support through Senior
Research Fellowship (Grant No. 09/093(0132)/2010).




\end{document}